\documentclass{article}
\usepackage{arxiv}

\usepackage[utf8]{inputenc} % allow utf-8 input
\usepackage[T1]{fontenc}    % use 8-bit T1 fonts
\usepackage[hidelinks]{hyperref}       % hyperlinks
\usepackage{url}            % simple URL typesetting
\usepackage{booktabs}       % professional-quality tables
\usepackage{amsfonts}       % blackboard math symbols
\usepackage{nicefrac}       % compact symbols for 1/2, etc.
\usepackage{microtype}      % microtypography
\usepackage{lipsum}		% Can be removed after putting your text content
\usepackage{graphicx}
\usepackage[square,numbers]{natbib}
\usepackage{doi}
\usepackage{caption}
\usepackage{subcaption}
\usepackage{amsmath}

\usepackage{bm}
\usepackage{url}
\usepackage[usenames,dvipsnames]{xcolor}

   \usepackage{stmaryrd}
\usepackage{amsmath}
\usepackage{amsfonts}
\usepackage{amssymb}
\usepackage{xparse}
\usepackage{cancel}
\usepackage[dvipsnames]{xcolor}
\usepackage{bm}
\usepackage{graphicx}
\usepackage{caption}
\usepackage{subcaption}
\usepackage{natbib}

\DeclareMathAlphabet{\mathbbold}{U}{bbold}{m}{n}

\newcommand{\mF}{\mathcal{F}}

\def\bmB{{\sf B}}

\def\bmD{{\sf D}}

\def\bmH{{\sf H}}

\def\bmQ{{\sf Q}}

\newcommand{\mean}[1]{\overline{#1}\,}

\newcommand{\meanp}[1]{\overline{#1}\,}

\newcommand{\logmeanp}[1]{\overline{#1}^{(\log)}\,}
\newcommand{\hlogmeanp}[1]{\overline{#1}^{(H\log)}\,}

\newcommand{\diffnopar}{\delta}
\newcommand{\diffpar}[1]{\diffnopar\!\left({#1}\right)}
\NewDocumentCommand\diff{ g }{
  \IfNoValueTF{#1}{\diffnopar}{\diffpar{#1}}
}
\newcommand{\diffcnopar}{\delta^{0}}
\newcommand{\diffcpar}[1]{\diffcnopar\!\left({#1}\right)}
\NewDocumentCommand\diffc{ g }{
  \IfNoValueTF{#1}{\diffcnopar}{\diffcpar{#1}}
 }

  \newcommand{\diffpnopar}{\delta^{+}}
\newcommand{\diffppar}[1]{\diffpnopar\!\left({#1}\right)}
\NewDocumentCommand\diffp{ g }{
  \IfNoValueTF{#1}{\diffpnopar}{\diffppar{#1}}
}
\newcommand{\diffmnopar}{\delta^{-}}
\newcommand{\diffmpar}[1]{\diffmnopar\!\left({#1}\right)}
\NewDocumentCommand\diffm{ g }{
  \IfNoValueTF{#1}{\diffmnopar}{\diffmpar{#1}}
}

\newcommand{\dps}{\delta_{\text{2pt}}}

\NewDocumentCommand\flux{ m g }{
  \IfNoValueTF{#2}{\mF_{{#1}}^{\,i+\frac{1}{2}}}{\mF_{{#1}}^{\,{#2}}}
}

\newcommand{\hmean}[1]{\overline{#1}^{H}}

\newcommand{\gmean}[1]{\overline{#1}^{G}}

\newcommand{\pmean}[1]{\overline{\overline{#1}}}

\newcommand{\CFL}{\text{CFL}}

\newtheorem{remark}{Remark}

\newcommand{\ECw}{EC\textsuperscript{W}}
\newcommand{\ECf}{EC\textsuperscript{F}}
\newcommand{\ECb}{EC\textsuperscript{B}}
\newcommand{\ECs}{EC\textsuperscript{S}}

\newcommand{\ESw}{ES\textsuperscript{W}}
\newcommand{\ESf}{ES\textsuperscript{F}}
\newcommand{\ESb}{ES\textsuperscript{B}}
\newcommand{\ESs}{ES\textsuperscript{S}}

   \graphicspath{{Figure/}}
   \usepackage{multirow}
\usepackage{makecell}
\usepackage{adjustbox}
   
   \usepackage{caption}
   \usepackage{subcaption}
   \usepackage{placeins}
   \usepackage[normalem]{ulem}
   \renewcommand{\d}{\mathrm{d}}
\title{Finite-difference compatible entropy-conserving schemes for the compressible Euler equations\thanks{Distribution Statement A: Approved for Public Release; Distribution is Unlimited. AFRL-2024-6076.}}

\author{ \href{https://orcid.org/0000-0002-6518-3114}{\includegraphics[scale=0.06]{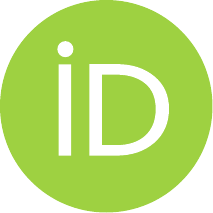}\hspace{1mm} Carlo {De~Michele}}\\
    Italian Aerospace Research Center (CIRA)\\
	Capua (CE), Italy \\
    Gran Sasso Science Institute (GSSI)\\
	L'Aquila, Italy \\
	\texttt{carlo.demichele@gssi.it} \\
	\And
	\href{https://orcid.org/0000-0002-7463-559X}{\includegraphics[scale=0.06]{orcid.pdf}\hspace{1mm}Ayaboe K.~Edoh} \\
    Amentum~--- Edwards Air Force Base\\
    CA 93524, USA\\
	\texttt{ayaboe.edoh.ctr@us.af.mil} \\
	\And
	\href{https://orcid.org/0000-0003-4943-9551}{\includegraphics[scale=0.06]{orcid.pdf}\hspace{1mm}Gennaro Coppola} \\
	Dipartimento di Ingegneria Industriale\\
	Universit\`a di Napoli ``Federico II''\\
	Napoli, Italy \\
	\texttt{gcoppola@unina.it} \\
}

\renewcommand{\shorttitle}{FD compatible EC schemes for the compressible Euler equations}%

\hypersetup{
pdftitle={Finite-difference compatible entropy-conserving schemes for the compressible Euler equations},
pdfsubject={physics.flu-dyn},
pdfauthor={Carlo De~Michele, Ayaboe K.~Edoh, Gennaro Coppola}
pdfkeywords={
compressible flow, finite difference, entropy conservation, summation-by-parts, logarithmic mean, kinetic-energy-preserving, pressure-equilibrium-preserving},
}

\begin{document}
\maketitle

\begin{abstract}
This paper introduces a family of entropy-conserving finite-difference discretizations for the compressible flow equations. 
In addition to conserving the primary quantities of mass, momentum, and total energy, the methods also preserve kinetic energy and pressure equilibrium.
The schemes are based on finite-difference (FD) representations of the logarithmic mean, establishing and leveraging a broader link between linear and nonlinear two-point averages and FD forms.
The schemes are locally conservative due to the summation-by-parts property and therefore admit a local flux form, making them applicable also in finite-volume and finite-element settings.
The effectiveness of these schemes is validated through various test cases (1D Sod shock tube, 1D density wave, 2D isentropic vortex, 3D Taylor--Green vortex) that demonstrate exact conservation of entropy along with conservation of the primary quantities and preservation of pressure equilibrium.
\end{abstract}

%% main text
\section{Introduction}

It is generally considered advantageous to discretize the compressible flow equations using supra-conservative~\cite{Veldman_SIAMRev_2021} schemes that are not only able to conserve the primary invariants, but also additional secondary ones.
Aside from the primary conservation of mass, momentum and total energy, structure-preserving methods for flows typically seek to improve discrete consistency with other secondary properties such as entropy, kinetic energy, pressure-equilibrium, angular momentum, helicity, enstrophy, etc.~\cite{Rozema_JT_2014,Kuya_JCP_2018,Shima_JCP_2021,Abgrall:2022b,Edoh_JCP_2022,Picklo:2025}.
In addition to improving the physical relevance and accuracy of the solution field, incorporating more consistencies can improve numerical stability, the benefits of which have been witnessed in increasingly complex physical settings such as for thermally perfect, multi-component, and reacting gases~\cite{Gouasmi:2020,Peyvan:2023,Ching:2024,Aiello_JCP_2025,Aiello_ArXiv_2025}.
For example, kinetic energy serves as a bounded $L_2$ estimate on the velocity variables for incompressible flows, while enstrophy serves a similar role with respect to a vorticity variable treatment~\cite{Arakawa_JCP_1966,Morinishi_JCP_1998}.
Such specialized quantities thus constitute mathematical entropies, which allow for a more rigorous statement on the robustness of the solution---even in the absence of natural stabilization mechanisms such as diffusion.
In other instances, methods have been designed to directly satisfy entropy dynamics in order to emphasize nonlinear stability and are then instead made to recover primary conservation of total energy as an auxiliary consequence~\cite{Ranocha_JSC_2018,Busto_SIAMJSC_2022,Abgrall2023}.
Overall, developing such structure-preserving methods is attractive for achieving a robust yet minimally dissipative method for use when calculating flows that would otherwise be sensitive to numerical damping, such as in Large Eddy Simulations of high-Reynolds number configurations~\cite{Volpiani_FTaC_2024}.

In incompressible flow, kinetic energy acts as a bounded $L_2$ measure of velocity, while enstrophy provides similar control in vorticity-based formulations~\cite{Arakawa_JCP_1966,Morinishi_JCP_1998}. These quantities function as mathematical entropies and can be used to assess nonlinear stability.
For compressible flows, kinetic energy no longer provides a bounded estimate on the flow field.
Instead, the thermodynamic entropy ($s$) is often used, as its conservation yields an energy-like statement with respect to a set of special transformation variables (i.e., the entropy variables)\footnote{A precise energy statement is available when considering conservation for the class of Harten entropies $h(s)$~\cite{Harten:1983}. However, unlike the thermodynamic entropy, these functions do not admit provable stability for Navier--Stokes configurations that include Fourier conduction terms.}; this notion is furthermore strongly tied to the symmetrization of the Euler equations~\cite{Merriam:1989a}. 
To this end, the current work introduces a new class of entropy-conserving methods for the compressible Euler system.
Notably, these schemes have the additional property of being compatible with finite differencing.

Entropy-conserving (EC) methods have traditionally been designed in terms of two-point numerical fluxes that are based on a jump condition constraint introduced by Tadmor~\cite{Tadmor_MC_1987}.
Their extension to high-order is then achieved via the notion of flux differencing~\cite{LeFloch:2002,Fisher_JCP_2013a}, wherein various low-order two-point fluxes are assembled in a type of Richardson extrapolation.
This overall flux framework naturally lends itself to the finite volume (FV) community and has also gained wide use for achieving entropy consistency with modern finite element (FE) schemes~\cite{Gassner:2013,Ranocha:2023}.
The congruence between flux (i.e., weak) forms and derivative (i.e., strong) forms is generally understood when discretizing linear functions with finite difference (FD) methods \cite{Pirozzoli_JCP_2010,Coppola_JCP_2019}.
A link between strong and weak forms, however, can also be established for nonlinearly stable discretizations in special instances.
For example, the entropy-conservative treatment of Burgers' equation can be expressed in a conservative FD form via a quadratic skew-symmetric splitting of the nonlinear term that induces the Heronian mean~\cite{Richtmeyer:1967,Jameson_JSC_2008}.
While discretely equivalent, the respective methods entail different algorithmic overheads in terms of their implementation, particularly when extending to high-order.
Such a direct duality between finite differencing and EC flux forms also exists for the incompressible flow equations \cite{Morinishi_JCP_1998}; however, the link has not yet been established in the context of the compressible Euler equations and may facilitate the discovery of new schemes.

Although FD formulations for the Euler equations can be designed to preserve linear invariants in addition to other properties like being kinetic-energy-preserving (KEP) and pressure-equilibrium-preserving (PEP), they traditionally do not exhibit the \emph{strict} EC property.
However, a notable class of FD splittings with heightened EC capabilities have been identified, which may be characterized as being \emph{quasi} entropy-conserving methods (qEC).
These include, among others, the recently developed kinetic-energy-entropy-preserving (KEEP) methods~\cite{Kuya_JCP_2018}.
The qEC schemes admit efficient FD representations and have been shown to greatly enhance numerical robustness across compressible regimes despite only being approximately entropy conserving.
Interesting to note, is that the flux form associated with these FD-qEC methods emulates the flux form's structure of Ranocha's EC scheme \cite{Ranocha:2018phd,Ranocha_CAMC_2021}\footnote{Ranocha's flux is equivalent to the earlier EC scheme of Chandrashekar \cite{Chandrashekar_CCP_2013} in terms of the convective treatment; however, the two methods differ in how the pressure terms are calculated, with Ranocha's rendition enabling pressure-equilibrium-preservation and also having a direct finite difference representation, which yields greater energy consistency with respect to the pressure-velocity coupling.}; the main point of contrast, however, is how the density ($\rho$) and internal energy ($e$) variables are averaged at the flux interface for the convective flux.
A variety of other qEC methods that feature this same structure have recently been proposed \cite{Tamaki_JCP_2022,DeMichele_JCP_2023,Kawai:2025} and may be termed asymptotically entropy-conserving; however, these forms are inherently flux-based in nature and are not traditionally associated with a finite difference splitting approach. 
While FD-qEC methods induce their flux averages (arithmetic or geometric) on the density and internal energy variables via quadratic and cubic splittings of the convective terms \cite{Blaisdell_ANM_1996,Kennedy_JCP_2008}, this framework does not readily admit the logarithmic-type means that are necessary for recovering strict conservation of the thermodynamic entropy in the case of perfect gases.
However, in the same way that standard FD discretizations (with central schemes) of split forms based on the divergence and advective forms of the convective terms have been linked to numerical fluxes based on bilinear and trilinear interpolations~\cite{Coppola_JCP_2019,DeMichele_JCP_2023}, it is of interest to find FD discretizations linked to a Tadmor-type EC method such as Ranocha's flux.

The current work develops new finite-difference compatible methods that are strictly EC in the context of a single-component ideal gas with the calorically perfect assumption.
In order to achieve this, the concept of split forms, which is traditionally based on the application of the product rule, is generalized by using the chain rule applied to suitable nonlinear transformations.
This is then used to generate different expressions for the discretized convective terms that are equivalent at the continuous limit. 
In this way, the intended nonlinear logarithmic flux averages for density and internal energy, $\tilde{\rho}$ and $\tilde{e}$, are first shown to be expressible as quotients of differences and then are 
linked to nonlinear numerical fluxes.
Rather than satisfying a Tadmor jump condition on the fluxes, the new methods are derived by requiring point-wise cancellation of the spurious volumetric terms that are induced in the discrete entropy equation that is associated with the primary system.
Enforcing these cancellations naturally reveals the necessary form of the special auxiliary variables; it also highlights the inherent relation between the choice in FD stencils and the derivative-based representation of the logarithmic means.
While it is possible to make FD methods \emph{globally} entropy conserving via correction procedures \cite{Abgrall2018,Coppola_JCP_2019,Abgrall:2022,Mantri:2024,Edoh2024,Alberti:2024,Alberti:2025} (or cell-wise for element-based methods), the current framework goes beyond by avoiding source terms in the resulting discrete entropy equation.
In addition to being EC, the new methods are made to be KEP and PEP as well.
Throughout, we employ finite differencing based on diagonal-norm summation-by-parts (SBP) operators \cite{DRF:2014,Zingg:2014}, including biased renditions \cite{Mattsson:2017,Ranocha:2024,Duru:2024}.
Notably, viable biased EC methods are identified for the first time---a feat that is facilitated by the current FD perspective.
A finite difference form of Ranocha's EC flux is then shown to be recovered as a special symmetric weighting of the two-point forward and backward biased operators.

The paper is organized as follows: Section~\ref{sec:2} provides preliminaries in terms of the governing equations and the SBP-based finite difference stencils; Section~\ref{sec:3} presents derivative-based representations of specialized nonlinear means; Section ~\ref{sec:4} introduces the new FD-EC methods; Section~\ref{sec:5} confirms the intended scheme properties relative to various test cases (1D density wave, 1D Sod shock tube, 2D isentropic vortex, 3D Taylor--Green vortex); and finally Section~\ref{sec:6} provides summary remarks and suggestions for further research.
Then in the Appendix, we furthermore supply the following: the entropy conservation proof in matrix-vector notation with SBP operators (Appendix \ref{sec:app_a}), the schemes expressed in Hadamard product form (Appendix \ref{sec:app_b}), an example of incorporating entropy stable artificial dissipation  (Appendix \ref{sec:app_c}), and employing the biased two-point fluxes in a multi-point fashion via flux differencing (Appendix \ref{sec:app_d}). 

%
%------
% 
\section{Preliminaries} \label{sec:2}

To better specify the context of our analysis, we start by considering the compressible Euler equations in $d$-dimensions, written as the balance equations for mass, momentum and total energy for a single-component gas in the absence of viscous and thermal effects
\begin{eqnarray}
\dfrac{\partial \rho}{\partial t} =&  - \dfrac{\partial  \rho u_n  }{\partial x_n} & = -\boldsymbol{\mathcal{C}}_\rho  \label{eq:Mass} \\
\dfrac{\partial \rho u_m}{\partial t} =& -\left(\dfrac{\partial  \rho u_n u_m  }{\partial x_n} + \dfrac{\partial  p  }{\partial x_m} \right) & = -(\boldsymbol{\mathcal{C}}_{\rho u_m} +  \boldsymbol{\mathcal{P}}_{\rho u_m})\label{eq:Momentum}  \\
\dfrac{\partial \rho E}{\partial t} =& - \left(\dfrac{\partial  \rho u_n E  }{\partial x_n} + \dfrac{\partial  p u_n  }{\partial x_n}\right) & = -(\boldsymbol{\mathcal{C}}_{\rho E} +  \boldsymbol{\mathcal{P}}_{\rho E}) \label{eq:TotEnergy} 
\end{eqnarray}
where $m$ and $n$ are integers ranging between $1$ and $d$ (the dimension) with repeated index notation being presumed.
In Eqs.~\eqref{eq:Mass}--\eqref{eq:TotEnergy}, $\rho$ is the density, $u_m$ is the Cartesian velocity in direction $m$, $p$ is the pressure, and $E$ is the total energy per unit mass which is made up of the sum of internal and kinetic energies ($E = e + \frac{1}{2}\sum_{m=1}^d u_m^2$). 
The equations are furthermore categorized in terms of the convective and pressure operators, $\boldsymbol{\mathcal{C}}$ and $\boldsymbol{\mathcal{P}}$ respectively. 
Kinetic energy consistency (i.e., that the combination mass and momentum induce a viable kinetic energy representation) and energy consistency (i.e., that the total energy is composed of the internal energy representation as well as the induced kinetic energy form) \cite{Coppola_JCP_2019,Kuya_JCP_2018} are motivated by the following additional analytic relations
\begin{eqnarray}
 \boldsymbol{\mathcal{P}}_{\rho E} &=&  \sum_{m=1}^d u_m\boldsymbol{\mathcal{P}}_{\rho u_m} + \boldsymbol{\mathcal{P}}_{\rho e} \nonumber \\
 &=& \sum_{m=1}^d \left(u_m \dfrac{\partial  p  }{\partial x_m} + p \dfrac{\partial  u_m  }{\partial x_m} \right) \label{eq:EnergyConst_P} 
 \\ \nonumber \\
\boldsymbol{\mathcal{C}}_{\rho E} &=&  \boldsymbol{\mathcal{C}}_{\rho k} + \boldsymbol{\mathcal{C}}_{\rho e}
 \quad \text{with} \
 \boldsymbol{\mathcal{C}}_{\rho k} = \sum_{m=1}^d\left[u_m\boldsymbol{\mathcal{C}}_{\rho u_m} - \frac{1}{2} u_m^2\boldsymbol{\mathcal{C}}_\rho\right] \nonumber \\
 &=& \dfrac{\partial  \rho u_n k  }{\partial x_n} + \dfrac{\partial  \rho u_n e  }{\partial x_n}  \quad \text{with} \ k = \frac{1}{2}\sum_{m=1}^d u_m^2.\label{eq:EnergyConst_C} 
\end{eqnarray}
Based on the additional physical constraints implied by the above, one is then left to determine the discrete representations of the following: $\boldsymbol{\mathcal{C}}_{\rho}$, $\boldsymbol{\mathcal{C}}_{\rho u_m}$, $\boldsymbol{\mathcal{C}}_{\rho e}$, 
$\boldsymbol{\mathcal{P}}_{\rho u_m}$, and $\boldsymbol{\mathcal{P}}_{\rho e}$.

%-----
\subsection{Entropy analysis}

We assume that pressure and internal energy are linked by the ideal gas equation of state, which implies $p = (\gamma-1)\rho e$, that the specific heats at constant pressure and volume and their ratio $\gamma = c_p/c_v$
are constant, and that $e = c_v T$ where $T$ is the absolute temperature. 
In this context, the thermodynamic entropy per unit mass is via Gibbs relation as 
\begin{equation} \label{eq:entropydef}
    \d s =  \frac{\d e}{T} -  R \frac{\d \rho}{\rho} \ \to \  s - s_\text{ref} = c_v \cdot \log (e/e_\text{ref}) - c_v(\gamma - 1) \cdot \log (\rho/\rho_\text{ref}) \ ,
\end{equation}
which presumes normalized quantities.
The entropy can then be shown to satisfy the conservation equation
\begin{equation}\label{eq:Entropy}
    \dfrac{\partial \rho s}{\partial t} = -\dfrac{\partial  \rho u_n s  }{\partial x_n}. 
\end{equation}
Eq.~\eqref{eq:Entropy}, which is valid in general for an arbitrary equation of state, is a statement that for non-viscous adiabatic flows the entropy of material particles remains unchanged, which in turn implies that global entropy within an Eulerian volume $\Omega$ does not change in time, except due to inflow or outflow at the boundaries. 
Eq.~\eqref{eq:Entropy} is a consequence of the system of Euler equations \eqref{eq:Mass}--\eqref{eq:TotEnergy} and of basic thermodynamics, and does not constitute an independent balance principle\footnote{Note that analogous conservation statements can be written for special functions of the entropy, $h(s)$, as proposed by Harten \cite{Harten:1983}. However, we focus on the thermodynamic entropy herein.}. 
As such, smooth solution fields satisfying Eqs.~\eqref{eq:Mass}--\eqref{eq:TotEnergy} also satisfy Eq.~\eqref{eq:Entropy}.
To see this, consider the set of $(2+d)$ entropy variables
\begin{eqnarray} \label{eq:entvars}
w \triangleq \frac{\partial \rho s}{\partial \underbrace{[\rho, \rho u_m, \rho E]}_{[q_1,\dots,q_{2+d}]}} = \frac{1}{T} \cdot \left[sT - e - \frac{p}{\rho} + \frac{1}{2}\sum_{m=1}^du_m^2,\ -u_m,\ 1\right] = \underbrace{[w_\rho, \ w_{\rho u_m}, \ w_{\rho E}]}_{[w_1, \dots ,w_{2+d}]}
\end{eqnarray}
and their contraction with the governing system:
\begin{eqnarray}
\dfrac{\partial \rho s}{\partial t} =  \sum_{m = 1}^{2+d} w_m \cdot \frac{\partial q_m}{\partial t } % &=& \nonumber \\
&=& w_\rho \cdot \frac{\partial \rho}{\partial t} + w_{\rho u_m} \cdot \frac{\partial \rho u_m}{\partial t} + w_{\rho E} \cdot \frac{\partial \rho E}{\partial t} \\
&=& \frac{1}{T} \left(sT  - e - \frac{p}{\rho} \right) \cdot \frac{\partial \rho}{\partial t} + \frac{1}{T} \cdot \frac{\partial \rho e}{\partial t} \nonumber \\
&=& s \cdot \frac{\partial \rho}{\partial t} + \rho \cdot \underbrace{\left(\frac{1}{T} \cdot \frac{\partial e}{\partial t} 
- \frac{p}{\rho^2 T} \cdot \frac{\partial \rho}{\partial t}\right)}_{\partial s / \partial t} . \nonumber \label{eq:EntVar_Euler_contraction}
\end{eqnarray}
The above manipulations invoke two key relations: 1) energy consistency (i.e., $\partial \rho e = \partial \rho E - (\sum_{m=1}^d u_m \partial \rho u_m - \frac{1}{2} u_m^2 \partial \rho) = \partial (\rho E - \frac{1}{2} \sum_{m=1}^d \partial \rho u_m^2)$ and 2) Gibbs relation (i.e., $T\partial s = \partial e  - (p/\rho^2) \partial \rho$).
When turning to discrete approximations, however, these types of links are typically lost in the sense that a consistent discretization of the system  \eqref{eq:Mass}--\eqref{eq:TotEnergy} does not imply that the discrete entropy is globally conserved. 
This behavior can be traced back to the failure of nominal operators to satisfy the usual rules of calculus that are valid for continuous operators---namely the product, chain, and summation-by-parts rules. EC schemes are designed, however, to exactly reproduce these structural properties of the continuous set of equations at the discrete level.

We will focus on the spatial discretization of the system of Eqs.~\eqref{eq:Mass}--\eqref{eq:TotEnergy} in the context of a semidiscretized procedure. 
This means that the analysis will assume that temporal integration can be carried out at the continuous level, whereas spatial approximation is obtained by using a discrete method.
The framework of a conservative Finite Difference method is adopted, where variables constitute the nodal values of the continuous unknown functions over a (typically uniform) Cartesian mesh. 
The formulation used is {\it conservative} in the sense that convective and derivative terms are approximated by numerical schemes that can be expressed as sum of differences of numerical fluxes at adjacent nodes along each Cartesian direction. 
This property discretely reproduces the divergence structure of the convective terms in Eqs.~\eqref{eq:Mass}--\eqref{eq:Entropy} and is hence referred to as the discretization's {\it local conservation} property. 
{\it Global conservation} (i.e., the property that the total amount of the balanced quantity is not affected by the convective terms, except for boundary contribution) is then guaranteed by the telescoping property, mimicking what happens at the continuous level.

For the sake of simplicity, but without loss of generality, we will expose our theory when possible with reference to the one-dimensional version of the system \eqref{eq:Mass}--\eqref{eq:TotEnergy}, such that $(m,n) = 1$.
The discretization of the space coordinate $x$ is made by considering a uniform mesh with nodal coordinates $x_i$ and width $(\Delta x)$. 
Although not described herein, the extension to non-uniform meshes can be accomplished by taking into account the variable sizes of the discretization into the time-derivative terms (\cite{Coppola_JCP_2023}), such as via the grid metrics Jacobian for curvilinear grids \cite{Zingg:2014}. 

%-----
\subsection{Conservative finite difference stencils}

In what follows, some difference operators will be used, and they are here defined.
First, consider a general finite difference stencil evaluated at node $i$
\begin{equation}
(\partial_x \phi)_i \approx (\delta \phi)_i = \frac{1}{\Delta x}\sum_k d_{i,i+k} \cdot \phi_{i+k} \ . \label{eq:FDstencil}
\end{equation}
In the above, we note that the FD stencil coefficients $d_{i,i+k}$ are dependent on the node at which the derivative is being estimated.
Herein, we assume that the progression of stencils is arranged such as to satisfy a Summation-by-Parts (SBP) property \cite{Zingg:2014, DRF:2014}, which is a discrete form of integration-by-parts, $\int_\Omega u\,\d v = \left(-\int_\Omega v\,\d u + (uv)|_{\partial \Omega}\right)$.
Considering a general quadratic decomposition of the input $\phi = uv$, then the SBP property for stencils $\delta^{\text c}$ associated with a centered operator is expressed as
\begin{align}
\sum_i h_i \cdot \ u_i (\delta^{\text c} v)_i &= -\sum_i h_i \cdot v_i (\delta^{\text c} u)_i + (u_r v_r - u_\ell v_\ell) \label{eq:SBP} 
\end{align}
where $h_i$ are scaling factors associated with a diagonal-norm of the SBP operator, which can be interpreted as a quadrature rule~\cite{Hicken:2013}, and $(\cdot)_{\ell/r}$ correspond to consistent approximations of the function at the left and right boundaries.
In the case of classic finite difference SBP operators featuring nodes at the boundary, then $(\cdot)_{\ell/r}$ is the nodal value at the boundary itself such that $u_\ell v_\ell = (uv)_\ell$; however, additional interpretations are possible via the generalized SBP formalism~\cite{DRF:2014} which accommodates the interpretation of such values as being consistent extrapolations or quasi-interpolations of the near boundary data.
Note that in the interior of the domain, away from boundaries, the $\delta^{\text c}$ stencils above are central and asymmetric.
The above SBP statement can be further extended to biased dual-pair SBP operators with local difference stencils $\delta^\pm$~\cite{Dovgilovich:2015,Mattsson:2017,Duru:2024} as
\begin{align}
\sum_i h_i \cdot \ u_i (\delta^\pm v)_i &= -\sum_i h_i \cdot v_i (\delta^\mp u)_i + (u_r v_r - u_\ell v_\ell) \label{eq:SBPpm}
\end{align}
with the relationship

\begin{equation}
\delta^\pm =  \overbrace{\frac{1}{2}(\delta^+ + \delta^-)}^{\delta^{\text c}} \ \pm \ \overbrace{\frac{1}{2}(\delta^+ - \delta^- )}^{\delta^{\text AD}} \ ,\label{eq:cent_dp_rel}
\end{equation}
where $\delta^{\text AD}$ constitutes an artificial dissipation (AD) term.
Rearranging terms in Eqs.~\eqref{eq:SBP} and \eqref{eq:SBPpm} reveals the quadratically split ``advection form" (i.e., $u \delta v + v \delta u$) that has been shown to yield a conservative telescoping-like representation~\cite{Pirozzoli_JCP_2010,Fisher_JCP_2013a,Coppola_JCP_2023}.
The nominal ``divergence form" (e.g., $\delta (uv)$) also features a conservative telescoping property.
In either case, general finite-volume fluxes can be derived from the SBP-compatible finite difference stencils via a recursive relation
\begin{align}
\mathcal{F}_{i+1/2} = \left\{
\begin{array}{l c l}
 h_i \cdot \ \delta^\pm (uv) + \mathcal{F}_{i-1/2} & \text{such that} & \mathcal{F}_{1/2} \equiv \mathcal{F}_\ell = (uv)_\ell  \\ \\
h_i \cdot \ \left( u_i (\delta^\pm v)_i + v_i (\delta^\mp u)_i \right) + \mathcal{F}_{i-1/2} & \text{such that} & \mathcal{F}_{1/2} \equiv \mathcal{F}_\ell = u_\ell v_\ell
\end{array}\right. \label{eq:FD-fluxrec}
\end{align}
where $\mathcal{F}_{i+1/2}$ are interface fluxes induced by the SBP discretization that live on a dual grid, $\bar{x}_{i+1/2}$, whose spacing is tied to the norm scaling such that $h_i = (\bar{x}_{i+1/2} - \bar{x}_{i-1/2})$ with $x_{1/2} \equiv x_\ell$~\cite{Fisher_JCP_2013b}.
In effect, global conservation along with an identifiable flux (anywhere in the domain, such as the boundary) is sufficient for guaranteeing a telescoping flux property throughout~\cite{Coppola_JCP_2023}.
While written in terms of the biased operators, Eq.~\eqref{eq:FD-fluxrec} also applies to the central operators per Eq.~\eqref{eq:cent_dp_rel}.

The stencil coefficients $d_{i,i+k}$ in Eq.~\eqref{eq:FDstencil} and their progression in space may be conveyed in the form of matrices $\bmD$ acting on a vector of nodal solution data.
For example, the standard second-order (two-point) finite difference SBP central operator and its dual decomposition are the following (assuming $(\Delta x) = 1$ for convenience),
\begin{align}
& (\text{two-point stencils}): \\ 
& \underbrace{\left[\begin{array}{r r r r r} 
\color{red} -1 & \color{red} 1 \\
-1/2 & 0 & 1/2 \\
& & \dots \\
& & -1/2 & 0 & 1/2 \\
& & & \color{red} -1 & \color{red}  1
\end{array}\right] =
\frac{1}{2} \cdot \left[\begin{array}{r r r r r} 
\color{red} -2 & \color{red} 2 \\
 & -1 & 1 \\
& & \dots \\
& &  & -1 & 1 \\
& & & \color{red} 0 & \color{red}  0
\end{array}\right] +
\frac{1}{2} \cdot \left[\begin{array}{r r r r r} 
\color{red} 0 & \color{red} 0 \\
-1 & 1  \\
& & \dots \\
& & -1 & 1 &  \\
& & & \color{red} -2 & \color{red}  2
\end{array}\right]
}_{\bmD^c = \frac{1}{2}\left(\bmD^+ + \bmD^- \right)} \label{eq:SBP-CD02} 
\end{align}
where the associated diagonal norm scaling is $\bmH = \text{diag}([\frac{1}{2},1,\dots,1,\frac{1}{2}])$~\cite{Ranocha:2024}.
Note that the red and black rows in Eq.~\eqref{eq:SBP-CD02} represent the boundary versus interior stencils, respectively\footnote{The central operator in \eqref{eq:SBP-CD02} is second order on the  interior and first order on the boundary. Meanwhile the biased operators feature a reduction in order, being first order on the interior and zeroth order (i.e., inconsistent) at the boundaries.}.
For this two-point operator, one can derive simple two-point fluxes responsible for calculating the interior interface fluxes $\mathcal{F}_{i+1/2}$ from the nodal data.
For the interior central stencil we have
\begin{equation}\label{eq:SplitFlux_DivAdv_Central}
    \begin{array}{r c l }
    \delta_{\text{2pt}} (uv) & \longrightarrow& \mathcal{F}(u,v)_{i+1/2,\text{div}} = \frac{(uv)_{i+1} + (uv)_i}{2} \triangleq \overline{uv} \\ \\
    u \delta_{\text{2pt}}  v + v \delta_{\text{2pt}}  u  & \longrightarrow&  \mathcal{F}(u,v)_{i+1/2,\text{adv}} = \frac{u_{i+1}v_i + u_i v_{i+1}}{2} \triangleq \pmean{(u,v)}
    \end{array}
\end{equation}
and for the interior biased stencils we have
\begin{equation}\label{eq:SplitFlux_DivAdv_Biased}
    \begin{array}{r c l}
    \delta^+_{\text{2pt}}  (uv) & \longrightarrow& \mathcal{F}^+(u,v)_{i,\text{div}} = (uv)_{i+ 1} \\ \\
    \delta^-_{\text{2pt}}  (uv) & \longrightarrow& \mathcal{F}^-(u,v)_{i,\text{div}} = (uv)_i \\ \\    
    u \delta^+_{\text{2pt}}  v + v \delta^-_{\text{2pt}}  u  & \longrightarrow&  \mathcal{F}^+(u,v)_{i,\text{adv}} = u_i v_{i+1} = \mathcal{F}^-(v,u)_{i,\text{adv}}
    \end{array}
\end{equation}
It is important to note that the interior advective form's flux for the two-point central operator is symmetric in its arguments; however, this is not the case for the biased splitting, which would provide additional flexibility in tailoring the behavior of the discretizations---for example with respect to interacting with sharp gradients in the solution field etc.
In special cases (e.g., interior nodes discretized by central stencils) such two-point fluxes can be combined to express the fluxes associated with high-order operators~\cite{Pirozzoli_JCP_2010,Fisher_JCP_2013b} in the fashion of Richardson-type extrapolations.
A convex combination of the divergence and advective forms is then possible and known to produce an independent averaging of the inputs.
For example, when these forms are evenly weighted, we have
\begin{equation}\label{eq:SplitFlux_Skw_CentrBias}
\begin{array}{r c l l}
\frac{1}{2} \left[\delta_{\text{2pt}}  (uv) + u \delta_{\text{2pt}}  v + v \delta_{\text{2pt}}  u \right] & \longrightarrow&  \frac{1}{2}\left[ \mathcal{F}(u,v)_{\text{div}} + \mathcal{F}(u,v)_{\text{adv}} \right]_{i+1/2} &= \frac{u_{i+1} + u_i}{2}\frac{v_{i+1} + v_i}{2} \\ \\
\frac{1}{2} \left[\delta^+_{\text{2pt}}  (uv) + u \delta^+_{\text{2pt}}  v + v \delta^-_{\text{2pt}}  u \right] & \longrightarrow& \frac{1}{2}\left[ \mathcal{F}^+(u,v)_{\text{div}} + \mathcal{F}^+(u,v)_{\text{adv}} \right]_{i+1/2} &= \frac{u_{i+1}+u_i}{2} v_{i+1} \\ \\ 
\frac{1}{2} \left[\delta^-_{\text{2pt}}  (uv) + u \delta^-_{\text{2pt}}  v + v \delta^+_{\text{2pt}}  u \right] & \longrightarrow& \frac{1}{2}\left[ \mathcal{F}^-(u,v)_{\text{div}} + \mathcal{F}^-(u,v)_{\text{adv}} \right]_{i+1/2} &= \frac{u_{i+1}+u_i}{2} v_{i} \\ \\ 
\frac{1}{2} \left[\delta^+_{\text{2pt}}  (uv) + u \delta^-_{\text{2pt}}  v + v \delta^+_{\text{2pt}}  u \right] & \longrightarrow& \frac{1}{2}\left[ \mathcal{F}^+(u,v)_{\text{div}} + \mathcal{F}^-(u,v)_{\text{adv}} \right]_{i+1/2} &= u_{i+1}\frac{v_{i+1} + v_i}{2} \\ \\ 
\frac{1}{2} \left[\delta^-_{\text{2pt}}  (uv) + u \delta^+_{\text{2pt}}  v + v \delta^-_{\text{2pt}}  u \right] & \longrightarrow& \frac{1}{2}\left[ \mathcal{F}^-(u,v)_{\text{div}} + \mathcal{F}^+(u,v)_{\text{adv}} \right]_{i+1/2} &= u_i \frac{v_i + v_{i+1}}{2} \\ \\ 
\end{array}
\end{equation}
Further combinations are also possible upon considering combinations of the central and biased schemes (not shown).
Such even weighting of the divergence and advective forms has been shown to be beneficial in terms of reducing nonlinear aliasing errors, at least in the context of central stencils~\cite{Blaisdell_ANM_1996,Moin:1997,Edoh:2020,Kuya:2022b}.

%
%------
\section{Specialized means via finite differences} \label{sec:3}

In the previous section, a correspondence has been mentioned between finite difference discretizations of derivatives of  products of two variables and numerical fluxes in the case of two-point discretizations. 
In particular, the divergence and advective forms with such central or non-symmetric difference operators have been associated with the class of two-point fluxes based on arithmetic averages.
In special instances, finite difference splitting can also be used to induce a two-point geometric average~\cite{Reiss_CF_2013,Edoh_JCP_2022}---for example, this happens when considering 
$\phi = (\phi^{1/2} \phi^{1/2})$, in which case $\mathcal{F}(\phi^{1/2},\phi^{1/2})_{i+1/2,\text{adv}}= \mathcal{F}^\pm(\phi^{1/2},\phi^{1/2})_{i+1/2,\text{adv}} = \sqrt{\phi_i \phi_{i+1}}$.
However, in many applications, more general fluxes comprised of nonlinear averages 
are necessary. 
These fluxes would seem to be outside the possible schemes that can be generated by classical finite difference discretizations. 
For example, with respect to designing EC schemes, one needs access to other nonlinear averages such as logarithmic means.
Yet, it has not been evident how to induce such expressions from a splitting technique as is possible with respect to the geometric mean.
It then seems natural to inquire if there is a generalized approach within the finite-difference framework to obtain the discrete equations coming from schemes based on nonlinear flux functions.
From Eqs.~\eqref{eq:SplitFlux_DivAdv_Central}--\eqref{eq:SplitFlux_Skw_CentrBias}, one observes that the application of the product rule for manipulating the convective term leads to various flux formulations that are based on arithmetic averages for the product of two variables.  
However, the product rule is a special case of the {\it chain} rule, which is not satisfied by discrete operators and is associated with the composition of arbitrary nonlinear functions other than the simple product of two variables. From this observation, it is argued that to obtain fluxes in which a nonlinear average arises (e.g., of the type of the logarithmic mean appearing in the Ranocha's EC mass flux~\cite{Ranocha:2018phd,Ranocha_CAMC_2021}: $\mF_{\rho} = \logmeanp{\rho}\mean{u}$ where $\logmeanp{\rho} = \left(\rho_{i+1}-\rho_{i}\right)/\left(\log\rho_{i+1}-\log\rho_{i}\right)$), nonlinear transformations based on the chain rule have to be leveraged to re-express the variables and/or their products. These transformations could then give different expressions that are equivalent on a continuous ground but that furnish different approximations when treated discretely. 
In these discrete forms, the peculiarities of the nonlinear transformation are expected to influence the form of the fluxes and induce new nonlinear averages, as motivated below.

The case of the logarithmic mean is exemplary and is easily treated by using the transformation 
\begin{equation}\label{eq:LogTransf}
    \rho = \left(\dfrac{1}{\rho}\right)^{-1} = 
    \left(\dfrac{\d\log\rho}{\d \rho}\right)^{-1} = 
    \dfrac{\d\rho}{\d\log\rho}
\end{equation}
where the last equality holds by virtue of the inverse function theorem applied to the bijective function $\log \rho$.
In Eq.~\eqref{eq:LogTransf} the nonlinear transformation expressing $\rho$ is specified through a derivative to make the chain rule relevant.
By using Eq.~\eqref{eq:LogTransf} in the evaluation of the convective term for the continuity equation, one has
\begin{equation}\label{eq:ConvTermLog}
    \frac{\partial \rho u}{\partial x} = \frac{\partial  }{\partial x}\left(\frac{\d \rho}{\d \log\rho} u\right).
\end{equation}
Although the equality in Eq.~\eqref{eq:ConvTermLog} is exact on a continuous ground, the discretization of the right-hand side will be associated with numerical fluxes that are different than those associated with the classical divergence form of the left-hand side. 
In fact, a coordinated discretization of the internal and external derivatives in Eq.~\eqref{eq:ConvTermLog} can be shown to be equivalent to a FV formulation based on the Ranocha's mass flux (cf.~Section~\ref{sec:4}).
Moreover, using split forms and  
various choices for the approximations of the derivative of $\rho$ furnishes different expressions for the fluxes, as for the split forms described in Eqs.~\eqref{eq:SplitFlux_DivAdv_Central}-\eqref{eq:SplitFlux_Skw_CentrBias}.
This type of procedure, in which FV schemes based on nonlinear fluxes are formulated as FD approximations, can be useful because a generalized treatment of the FD formulations can generate new potentially useful discretizations. An example is studied in Section~\ref{sec:4} with reference to EC formulations based on the logarithmic mean.

A generalization of this procedure can be attempted by using a class of transformations associated with more general nonlinear averages, among which logarithmic and arithmetic means are particular cases.
A sufficiently general transformation based on the use of derivatives to express a variable $\phi$ is given by the following relation, valid for positive values of $\phi$: 
\begin{eqnarray}\label{eq:NonlinTrasformation}
    \phi &=& \dfrac{\beta}{\beta+1}\dfrac{\d \phi^{\beta+1}}{\d \phi^{\beta}} \\ 
    &=& \dfrac{\beta}{\beta+1}\left[\dfrac{\d \phi^{\beta+1}}{\d \phi}\dfrac{\d \phi}{\d \phi^{\beta}}\right] = \dfrac{\beta}{\beta+1}\left[\dfrac{\d \phi^{\beta+1}}{\d \phi}\left(\dfrac{\d \phi^{\beta}}{\d \phi}\right)^{-1}\right] \nonumber \\
&=&  \dfrac{\beta}{\beta+1}\left[\left(\beta+1\right)\phi^{\beta}\left(\beta \phi^{\beta-1}\right)^{-1}\right] = \dfrac{\beta}{\beta+1}\left[\left(\dfrac{\beta+1}{\beta}\right)\dfrac{\phi^{\beta}}{\phi^{\beta-1}}\right] = \phi. \nonumber
\end{eqnarray}
The validity of Eq.~\eqref{eq:NonlinTrasformation} is easily verified by observing that the function $\phi^{\beta}$ is a strictly monotone function for positive values of $\phi$. 
Eq.~\eqref{eq:NonlinTrasformation} represents the type of fundamental nonlinear transformations used herein to obtain the generalized FD schemes corresponding with more general nonlinear fluxes.
Further note that when $\beta = 1$, then $\phi$ above may take all real values, both negative and positive. 
An important observation is that a straightforward evaluation of Eq.~\eqref{eq:NonlinTrasformation} breaks down for the cases $\beta=0$ and $\beta=-1$.
In such instances, we take the respective limit and employ logarithmic substitutions
\begin{eqnarray}
    && \lim_{\beta\rightarrow 0} \ \dfrac{\beta}{\beta+1}\left[\left(\dfrac{\beta+1}{\beta}\right)\dfrac{\phi^{\beta}}{\phi^{\beta-1}}\right] = 
    \dfrac{1}{\phi^{-1}} = \left(\dfrac{\d  \log\phi}{\d \phi}\right)^{-1} = \dfrac{\d  \phi}{\d  \log\phi} \label{eq:logmean_lim} \\ \nonumber \\
    && \lim_{\beta\rightarrow -1} \dfrac{\beta}{\beta+1}\left[\left(\dfrac{\beta+1}{\beta}\right)\dfrac{\phi^{\beta}}{\phi^{\beta-1}}\right] = 
    \phi^{-1}/\phi^{-2} = \left(-\dfrac{\d \log(1/\phi)}{\d (\phi)}\right)\left(-\dfrac{\d  \phi}{\d  1/\phi}\right) = \dfrac{\d \log(1/\phi)}{\d (1/\phi)}  \label{eq:hlogmean_lim}
\end{eqnarray}
recovering the type of transformation illustrated with reference to the Ranocha's mass and internal energy flux.

Equation~\eqref{eq:NonlinTrasformation} therefore re-expresses $\phi$ in terms of a new parameterized function comprised of its derivatives, $\phi = f_{\overline{\phi}}(\beta)$.

Substituting finite difference approximations yields a consistent discrete filtered quantity
\begin{equation}
f_{\overline{\phi}}(\beta) \triangleq 
\dfrac{\beta}{\beta+1}\dfrac{\delta  \phi^{\beta+1}}{\delta \phi^{\beta}} \approx \phi \ ,
\end{equation}
with the limiting case of two-point derivatives recovering various discrete mean values:
\begin{enumerate}
    \item $\beta = 1 \quad\rightarrow\quad${Arithmetic mean}:\\ 
    $f_{\overline{\phi}}\left(1\right) =  \dfrac{1}{2}\dfrac{\delta\phi^2}{\delta\phi} \ \to \ \dfrac{1}{2}\dfrac{\dps \phi^2}{\dps \phi} = \dfrac{1}{2}\dfrac{\phi_{a}^2-\phi_{b}^2}{\phi_{a}-\phi_{b}} = 
    \dfrac{\phi_{a}+\phi_{b}}{2} = \mean{\phi}$ \\

       \item $\beta = 0 \quad\rightarrow\quad${Logarithmic mean}: \\    $f_{\overline{\phi}}\left(0\right) =  \dfrac{\delta \phi}{\delta \log\phi} \ \to \ \dfrac{\dps \phi}{\dps \log\phi}= \dfrac{\phi_{a}-\phi_b}{\log\phi_{a}-\log\phi_b} =
    \logmeanp{\phi}$ \\    
    \item $\beta = -\dfrac{1}{2} \quad\rightarrow\quad${Geometric mean}: \\     $f_{\overline{\phi}}\left(-1/2\right) =  -\dfrac{\delta\sqrt{\phi}}{\delta (1/\sqrt{\phi})} \ \to \ \dfrac{\dps\sqrt{\phi}}{\dps (1/\sqrt{\phi})}  =  \dfrac{\sqrt{\phi_{a}}-\sqrt{\phi_{b}}}{\dfrac{1}{\sqrt{\phi_{a}}}-\dfrac{1}{\sqrt{\phi_{b}}}}  = \sqrt{\phi_{a}\phi_b} = \gmean{\phi}$ \\
    \item $\beta = -1 \quad\rightarrow\quad${``Harmonic-logarithmic" mean}:  \\    $f_{\overline{\phi}}\left(-1\right) =  \dfrac{\delta\log(1/\phi)}{\delta(1/\phi)} \ \to \ \dfrac{\dps\log(1/\phi)}{\dps(1/\phi)}= \dfrac{\log (1/\phi_{a}) -\log (1\phi_b)}{\dfrac{1}{\phi_{a}}-\dfrac{1}{\phi_b}} =
    \hlogmeanp{\phi}$ \\
    \item $\beta = -2 \quad\rightarrow\quad${Harmonic mean}: \\     $f_{\overline{\phi}}\left(-2\right) =  \dfrac{\delta \phi^{-1}}{\delta \phi^{-2}} \ \to \ \dfrac{\dps \phi^{-1}}{\dps \phi^{-2}}= \dfrac{\dfrac{1}{\phi_{a}}-\dfrac{1}{\phi_{b}}}{\dfrac{1}{\phi_{a}^2}-\dfrac{1}{\phi_{b}^2}}= 
    2 \dfrac{\phi_b\phi_{a}}{\phi_b+\phi_{a}}=
    \hmean{\phi}$        
    \end{enumerate}
Other means may be recovered by the above formulas, such as the the Heronian mean when $\beta = 1/2$.
Also note that in the case of two-point means, we have the property that $f_{\overline{\phi}}(\beta_a) > f_{\overline{\phi}}(\beta_b) $ when 
$\beta_a > \beta_b$~\cite{Chen_MM_2005}.
Also, specialized procedures for dealing with the risk of a singular denominator will furthermore be necessary in the case of the logarithmic-based means (e.g., Ismail/Roe's procedure~\cite{Ismail_JCP_2009}).

    Equation~\eqref{eq:NonlinTrasformation} thus presents a general parameterized family of filters based on derivatives, where a mean of the inputs is recovered in the special case of two-point stencils.
    To see this latter property, the two-point discretization of $f_{\bar{\phi}}(\beta)$ can be re-written in integral form\footnote{Another well-known class of means based on differences are the Stolarsky means~\cite{Stolarsky_MM_1975,Stolarsky_AMM_1980}, which also coincide with some of the popular averages mentioned above (e.g., arithmetic, geometric, logarithmic). 
The Stolarski means induce an integral form, $\overline{\phi^{\beta}} \cdot \int_1^2  \,\d \phi = \int_1^2 \phi^\beta \,\d\phi$ but with the additional approximation that $\overline{\phi^\beta} \approx \bar{\phi}^\beta$.} as
\begin{equation}
\bar{\phi} \cdot \frac{1}{\beta} \cdot \delta_\text{2pt} \phi^{\beta} = \frac{1}{\beta +1} \cdot \delta_\text{2pt}  \phi^{\beta+1} \quad
\to \quad \bar{\phi} \cdot  \int_1^2 \phi^{\beta -1} \,\d\phi = \int_1^2 \phi^{\beta} \,\d\phi \ .
\end{equation}
Thus we see that $\bar{\phi}$ results from the use of the intermediate value theorem and therefore should lie within the interval $[\phi_1,\phi_2]$.
We note that the above recovers the same two-point means proposed in~\cite{Chen_MM_2005} 
\footnote{While the present work independently developed the current derivative-based representation, we acknowledge the work of~\cite{Chen_MM_2005} towards the generalization provided in Eq.~\eqref{eq:NonlinTrasformation}.}; however, the differential perspective offered by Eq.~\eqref{eq:NonlinTrasformation} directly generalizes to difference stencils of arbitrary length (e.g., high-order differencing), albeit coming at the loss of guaranteeing that the resulting filtered variable is bounded by its inputs.

The following section leverages the ability to explicitly generate nonlinear mean quantities such as the ones presented above in order to develop new entropy consistent methods that are finite-difference compatible. 
Specifically, we consider the following re-interpretation of the density and internal energy thermodynamic variables as part of re-writing their respective convective terms,
\begin{eqnarray}
    \left.\begin{array}{l}
    \rho = f_{\overline{\rho}}\left(\beta = 0\right) 
    = \frac{\d \rho}{\d \log\rho} \\ \\
            e = f_{\overline{e}}\left(\beta = -1\right) 
        = \frac{\d \log e^{-1}}{\d e^{-1}}
        \end{array}\right\}
    & \to & 
    \left\{\begin{array}{l}
    \frac{\partial  \rho u}{\partial  x} = \frac{\partial   }{\partial  x}\left(\frac{\d \rho}{\d \log\rho} u\right) \\ \\
        \frac{\partial  \rho u e}{\partial  x} = \frac{\partial   }{\partial  x}\left(\frac{\d \rho}{\d \log\rho}\,u\,\frac{\d \log e^{-1}}{\d e^{-1}}\right)
        \end{array}\right.
\end{eqnarray}
As will be shown in the following section, a direct FD treatment of these analytic forms results in discrete entropy conservation for the current calorically perfect setting with entropy defined in Eq.~\eqref{eq:entropydef}.

%------
%
\section{A class of finite-difference entropy-conserving methods} \label{sec:4}

In order to devise the FD-EC schemes, we first consider the entropy equation that results from contracting the entropy variables in Eq.~\eqref{eq:entvars} with the governing Euler system:
\begin{eqnarray}
   \boldsymbol{\mathcal{C}}_{\rho s} &=& 
   w_\rho \cdot \boldsymbol{\mathcal{C}}_{\rho} + \sum_{m=1}^d w_{\rho u_m} \cdot \left(\boldsymbol{\mathcal{C}}_{\rho u_m} + \boldsymbol{\mathcal{P}}_{\rho u_m} \right) + w_{\rho E} \cdot \left(\boldsymbol{\mathcal{C}}_{\rho E} + \boldsymbol{\mathcal{P}}_{\rho E} \right) \nonumber \\
   &=& \left[s - \gamma c_v \right] \cdot \boldsymbol{\mathcal{C}}_{\rho} + \frac{1}{T} \cdot \left(\boldsymbol{\mathcal{C}}_{\rho e} + \boldsymbol{\mathcal{P}}_{\rho e} \right) % \nonumber \\
\end{eqnarray}
Note that the above leverages the energy consistent relations from Eq.~\eqref{eq:EnergyConst_P} and \eqref{eq:EnergyConst_C}, namely that the discretization of total energy is comprised of an internal energy discretization plus a kinetic energy component that stems directly from the discretized mass and momentum equations.
The current intent is then to recover the entropy conservation statement by judiciously discretizing the primary system.
To this end, we first assume a direct discretization of the respective convective and pressure terms previously given in Eqs.~\eqref{eq:Mass}--\eqref{eq:EnergyConst_C}.
The centered and biased FD stencil renditions for each term associated with the $(n)$ direction derivative are given in the following Tables \ref{table:1} and \ref{table:2}, where the $\boldsymbol{\mathcal{C}}$ and $\boldsymbol{\mathcal{P}}$ symbols can now refer to discrete representations based on the context.
We allow for additional flexibility in how the density and internal energy will be defined via the specialized FD means presented in Section~\ref{sec:3}; the necessary forms for achieving exact entropy conservation are derived below.
Note that the choice of $\boldsymbol{\mathcal{C}}_{\rho}$ automatically specifies the momentum treatment $\boldsymbol{\mathcal{C}}_{\rho u}$ that is necessary for fulfilling kinetic-energy-preservation (KEP).
As we presume a divergence-like discretization of mass, the associated KEP form is that of Feiereisen~\cite{Feiereisen_1981}.
Meanwhile the internal energy convective term is also in divergence form and therefore will naturally admit pressure-equilibrium preservation.
These scheme properties are further discussed in Section~\ref{sec:4p1}.
Note that while the split forms presented in the tables are written relative to the local stencil, a matrix-vector rendition stems naturally, as utilized in Appendix \ref{sec:app_a} (e.g., $a\delta_x b \to [\text{diag}\{{\bf a}\}] {\bmD}_x{\bf b}$). 
Alternatively, Appendix \ref{sec:app_b} provides the methods in a Hadamard product form.

\begin{table}[h!]
\center 
\begin{tabular}{l | l   }
Centered &   \\ \hline   \\
$\boldsymbol{\mathcal{C}}_{\rho}^{(n)}$ & $\delta_{x_n}^{\text c} (\tilde{\rho} u_n)$
 \\ \\
$\boldsymbol{\mathcal{C}}_{\rho u_m}^{(n)}$ & $ \frac{1}{2}\left[\delta_{x_n}^{\text c} (\tilde{\rho} u_n u_m) + \tilde{\rho} u_n \delta_{x_n}^{\text c} u_m + u_m \delta_{x_n}^{\text c} (\tilde{\rho} u_n) \right]$  \\ \\
$\boldsymbol{\mathcal{C}}_{\rho e}^{(n)}$ & $\delta_{x_n}^{\text c} (\tilde{\rho} u_n \tilde{e})$  \\ \\
$\boldsymbol{\mathcal{P}}_{\rho u_m}^{(n)}$ & $ \delta_{x_m}^{\text c} p$ if $n=m$, otherwise $0$  \\ \\
$\boldsymbol{\mathcal{P}}_{\rho e}^{(n)}$ & $  p \delta_{x_n}^{\text c} u_n$  \\ \\
\end{tabular}
\caption{A kinetic-energy and pressure-equilibrium preserving finite-difference discretization of Euler derivatives assuming SBP-compatible stencils. In the case of the FD-EC methods with central convective operators, $\boldsymbol{\mathcal{C}}^{(n)}$, the auxiliary variables are 
$\tilde{\rho} \equiv \frac{\delta_{x_n}^{\text c} \rho}{\delta_{x_n}^{\text c} \log \rho} \triangleq \overline{\rho}^{(\log,n)}$ and 
$\tilde{e} \equiv \frac{\delta_{x_n}^{\text c} \log e^{-1}}{\delta_{x_n}^{\text c} e^{-1}} \triangleq \overline{e}^{(H\log,n)}$.} \label{table:1}
\end{table}

\begin{table}[h!]
\center 
\begin{tabular}{l | l }
 Biased &  \\ \hline   \\
$\boldsymbol{\mathcal{C}}_{\rho}^{(n,\pm)} $ & $ \delta^\pm_{x_n} (\tilde{\rho} u_n)$ \\ \\
$\boldsymbol{\mathcal{C}}_{\rho u_m}^{(n,\pm)}$& $ \frac{1}{2}\left[\delta_{x_n}^\pm (\tilde{\rho} u_n u_m) + \tilde{\rho} u_n \delta_{x_n}^\mp u_m + u_m \delta_{x_n}^\pm (\tilde{\rho} u_n) \right]$ \\ \\
$\boldsymbol{\mathcal{C}}_{\rho e}^{(n,\pm)}$ & $ \delta^\pm_{x_m} (\tilde{\rho} u_n \tilde{e})$  \\ \\
$\boldsymbol{\mathcal{P}}_{\rho u_m}^{(n,\pm)}$ & $  \delta^\mp_{x_m} p$ if $n=m$, otherwise $0$  \\ \\
$\boldsymbol{\mathcal{P}}_{\rho e}^{(n,\pm)}$ & $ p\delta^\pm_{x_n} u_n$ \\ \\
\end{tabular}
\caption{A kinetic-energy and pressure-equilibrium preserving finite-difference discretization of Euler derivatives assuming SBP-compatible stencils. In the case of the FD-EC methods with biased convective operators, $\boldsymbol{\mathcal{C}}^{(n,\pm)}$, the auxiliary variables are 
$\tilde{\rho} \equiv \frac{\delta_{x_n}^\mp \rho}{\delta_{x_n}^\mp \log \rho} \triangleq \overline{\rho}^{(\log,n,\mp)}$ and 
$\tilde{e} \equiv \frac{\delta^\mp \log e^{-1}}{\delta^\mp e^{-1}} \triangleq \overline{e}^{(H\log,n,\mp)}$.} \label{table:2}
\end{table}

The additional entropy conservative property is enforced by choosing $\tilde{\rho}$ and $\tilde{e}$ such that any spurious volumetric terms are canceled and all that remain are consistent boundary terms. 
This is required in a point-wise manner in order to further guarantee \emph{local} conservation in the discrete entropy dynamics.
To see this, one employs the presumed form of the discretized convective and pressure terms while furthermore employing the calorically perfect gas assumptions with respect to the definition of entropy, as well as re-writing $1/T = c_v/e$; these manipulations yield the following (presented in one-dimension and the right-biased velocity rendition for simplicity):
\begin{eqnarray}
    \boldsymbol{\mathcal{C}}_{\rho s} \big|_{\partial \Omega} &=& \int_\Omega \boldsymbol{\mathcal{C}}_{\rho s} \nonumber \\  &=& \int_\Omega \left[\left(c_v \cdot \log \frac{e}{e_\text{ref}}- c_v(\gamma - 1) \cdot \log \frac{\rho}{\rho_\text{ref}} - \gamma c_v \right) \cdot \boldsymbol{\mathcal{C}}_{\rho} + \frac{c_v}{e} \cdot \left(\boldsymbol{\mathcal{C}}_{\rho e} + \boldsymbol{\mathcal{P}}_{\rho e} \right) \right] \nonumber \\
       &=& \underbrace{c_v \int_\Omega \left[ \frac{1}{e} \cdot  \boldsymbol{\mathcal{P}}_{\rho e} - (\gamma-1) \log \frac{\rho}{\rho_\text{ref}} \cdot \boldsymbol{\mathcal{C}}_\rho\right]}_{\color{blue}(1)} 
 + \underbrace{c_v\int_\Omega \left[ \left(\log \frac{e}{e_\text{ref}} - \gamma \right) \cdot \boldsymbol{\mathcal{C}}_\rho + \frac{1}{e} \cdot \boldsymbol{\mathcal{C}}_{\rho e}\right]}_{\color{red}(2)}  \nonumber \\  \label{eq:EntropyIntegral} 
\end{eqnarray}
The integral notation used actually implies summation across nodal points in the discrete setting according to the SBP diagonal norm that encodes the interface flux spacings~\cite{Fisher_JCP_2013b}.
In terms of variable units, the explicit normalization of internal energy and density by their reference quantities within the logarithmic expressions of entropy have now been included for clarity of presentation.
As an example, we show this below in Eqs.~\eqref{eq:FD-ECTerm1} and \eqref{eq:FD-ECTerm2} for the one-dimensional case assuming a set of biased operators (note that the same mechanics apply for the opposite biasing and for the central stencils as well).
Appendix \ref{sec:app_a} shows the analogous entropy-conservation proofs with SBP operators and matrix-vector notation.

\begin{eqnarray}
\underbrace{c_v \int_\Omega \left[ \frac{1}{e} \cdot  \boldsymbol{\mathcal{P}}^{(+)}_{\rho e} - (\gamma-1) \log \frac{\rho}{\rho_\text{ref}} \cdot \boldsymbol{\mathcal{C}}^{(+)}_\rho\right]}_{\color{blue}(1)}: 
&&c_v \int_\Omega \left[ e^{-1} p \cdot \delta_x^+ u - (\gamma-1) \log \frac{\rho}{\rho_\text{ref}} \cdot \delta_x^+ (\tilde{\rho} u)  \right] \nonumber \\
&&= c_v(\gamma-1) \int_\Omega \left[ \rho \cdot \delta_x^+ u -  \log \frac{\rho}{\rho_\text{ref}} \cdot \delta_x^+ (\tilde{\rho} u)  \right] \nonumber\\
&& = c_v(\gamma-1) \left[ \rho u - \log \frac{\rho}{\rho_\text{ref}} \cdot \tilde{\rho} u \right] \big|_{\partial \Omega} \nonumber \\
&& - c_v(\gamma-1) \int_\Omega u \underbrace{\left[ \delta_x^- \rho -  \tilde{\rho} \cdot  \delta_x^-\left( \log \frac{\rho}{\rho_\text{ref}} \right) \right]}_{\text{need} \ =0} \label{eq:FD-ECTerm1} \\ \nonumber \\
\underbrace{c_v\int_\Omega \left[ \left(\log \frac{e}{e_\text{ref}} - \gamma \right) \cdot \boldsymbol{\mathcal{C}}^{(+)}_\rho + \frac{1}{e} \cdot \boldsymbol{\mathcal{C}}^{(+)}_{\rho e}\right]}_{\color{red}(2)}: 
&& c_v\int_\Omega \left[ \left(\log \frac{e}{e_\text{ref}} - \gamma \right) \cdot \delta_x^+ (\tilde{\rho} u )  + e^{-1} \cdot \delta_x^+ (\tilde{\rho} u \tilde{e}) \right] \nonumber\\
&& = c_v \left[ \left(\log \frac{e}{e_\text{ref}} - \gamma \right) \cdot  (\tilde{\rho} u )  + e^{-1} \cdot (\tilde{\rho} u \tilde{e}) \right] \big|_{\partial \Omega} \nonumber\\
&& - c_v\int_\Omega \tilde{\rho} u \underbrace{\left[ -\delta_x^- \left( \log \frac{e^{-1}}{e_\text{ref}^{-1}} \right)  +   \tilde{e} \cdot \delta_x^- e^{-1} \right]}_{\text{need} \ = 0} \label{eq:FD-ECTerm2}
\end{eqnarray}
The exposition above in Eqs.~\eqref{eq:FD-ECTerm1} and \eqref{eq:FD-ECTerm2} places no stipulations on the order of the difference operators and only requires the SBP property and the use of dual operators in the case of biased stencils.
Also, while the current example is shown for a single dimension, it should be understood that the set of specialized mean quantities $\tilde{\rho}$ and $\tilde{e}$ will need to be calculated for each directional derivative (see Remark~\ref{remark:3}).

The boundary surface estimates are based on the extrapolation formulas associated with the SBP operators per Eq.~\eqref{eq:SBP} and \eqref{eq:SBPpm}.
From the SBP manipulations in Eqs.~\eqref{eq:FD-ECTerm1} and \eqref{eq:FD-ECTerm2}, one can identify the consistent entropy boundary terms that are induced by the resulting FD scheme.
For example, with respect to the left boundary one can re-arrange terms to recover the associated high-order consistent entropy flux
\begin{align}
    \boldsymbol{\mathcal{C}}_{\rho s} \big|_\ell \equiv \mathcal{F}_{\rho s,\ell} 
    = \overbrace{(\tilde{\rho} u)_\ell \cdot  \left(c_v \left(\log \frac{e}{e_\text{ref}}\right)_\ell - c_v(\gamma-1) \left( \log \frac{\rho}{\rho_\text{ref}}\right)_\ell\right) }^{(\tilde{\rho} u)_\ell s_\ell \ \sim \ (\rho u s)_\ell } \label{eq:entropybcflux}
    %\\    
    %
%
    % && 
    +  \underbrace{\gamma c_v \left(\rho_\ell u_\ell - (\tilde{\rho} u )_\ell \right) - c_v \left(\rho_\ell u_\ell - (\tilde{\rho} u \tilde{e})_\ell  (e^{-1})_\ell \right)}_{O(\Delta x^p)} \nonumber \\
%
    %\nonumber
\end{align}
Equation~\eqref{eq:entropybcflux} is valid for all of the SBP stencils---whether central or forward/backward biased.
We further note that in cases where $\tilde{\rho} \ne \rho$ and $\tilde{e} \ne e$ (as will be the case for the resulting EC schemes) then the resulting boundary flux includes a persistent zero-consistent component. 
The spurious entropy volumetric terms that result from the summation-by-parts manipulations in Eqs.~\eqref{eq:FD-ECTerm1} and \eqref{eq:FD-ECTerm2} are cancelled \emph{point-wise} by selecting the auxiliary variables $\tilde{\rho} \equiv \frac{\delta_x^- \rho}{\delta_x^- \log \rho} \triangleq \overline{\rho}^{(\log,-)}$ and $\tilde{e} \equiv \frac{\delta_x^- \log e^{-1}}{\delta_x^- e^{-1}} \triangleq \overline{e}^{(H\log,-)}$.
Per the discussion of Section~\ref{sec:3}, these new variables constitute specialized averages of the density and internal energy.
The resulting method is then successfully entropy conserving and can also be associated with a local entropy-conserving Tadmor-type flux condition (see Remark~\ref{remark:2}).
The finite-difference representation that is induced in the discrete entropy equation can be identified from the structure of the boundary flux in Eq.~\eqref{eq:entropybcflux}.
Specifically, the forward biased rendition gives
\begin{eqnarray}
\boldsymbol{\mathcal{C}}_{\rho s}^{(+)} &=& s \delta_x^+ (\tilde{\rho}u) + \tilde{\rho}u \delta_x^- s \\
&& \ + \underbrace{\left[c_v(\gamma-1) \cdot \left(\rho \delta^+_x u + u \delta^-_x \rho \right) - \gamma c_v \cdot  \delta_x^+ (\tilde{\rho}u) + c_v \left( e^{-1}\delta_x^+ (\tilde{\rho}u \tilde{e}) + \tilde{\rho}u \tilde{e} \delta_x^- e^{-1}\right) \right]}_{O (\Delta x^p)} \ .\nonumber
\end{eqnarray}
Thus we recover a split form in the entropy dynamics that is based on the special nonlinear averages $\tilde{\rho}$ and $\tilde{e}$
Besides the main entropy terms, there are also zero-consistent and telescoping (i.e., conservative) corrective contributions.
Note that analogous expressions arise when switching the biasing and also when using the central operators (not shown).

\begin{remark}\label{remark:1}
    The groupings identified in Eq.~\eqref{eq:EntropyIntegral} for achieving point-wise cancellations of the spurious entropy dynamics are non-unique and can lead to different EC methods. This flexibility in scheme design is an advantageous characteristic of the proposed FD-EC methodology. The current choice of groupings results in  definitions for $\tilde{\rho}$ and $\tilde{e}$ that are solely functions of density and internal energy, respectively.
\end{remark}

\begin{remark} \label{remark:2}
The point-wise cancellations of the spurious dynamics in the discrete entropy equation (see Eqs.~\eqref{eq:FD-ECTerm1}--\eqref{eq:FD-ECTerm2}) suggest that the FD-EC schemes also satisfy the following Tadmor-type EC flux condition: $\sum_{m=1}^{2+d}(w_{m,i+1} - w_{m,i}) \mathcal{F}^{(n)}_{m,i+1/2} = (\tilde{\psi}^{(n)}_{i+1} - \tilde{\psi}^{(n)}_i)$ for $i = 1,\dots,N-1$,  where $\tilde{\psi}^{(n)}_1 \equiv \tilde{\psi}^{(n)}_\ell = (\sum_{m}^{2+d} w_{m,1}\mathcal{F}^{(n)}_{m,\ell} - \mathcal{F}^{(n)}_{\rho s,\ell})$ (see Section~3.2.1 in \cite{Fisher_JCP_2013b}).
\end{remark}

\begin{remark} \label{remark:3}
    The specialized auxiliary variables $\tilde{\rho}$ and $\tilde{e}$ need to be calculated with respect to each directional derivative. For example, fluxes in the $n$ direction would require calculating $\widetilde{\rho} \triangleq \frac{\delta_{x_n}^\pm \rho}{\delta_{x_n}^\pm \log \rho} \equiv \overline{\rho}^{(\log,n,\pm)}$ and $\widetilde{e} \triangleq \frac{\delta_{x_n}^\pm \log e^{-1}}{\delta_{x_n}^\pm e^{-1} } \equiv \overline{e}^{(H\log,n,\pm)}$ in the biased case. 
\end{remark}

The resulting FD-EC scheme is conservative and induces a local flux form.
The recursive procedure from Eq.~\eqref{eq:FD-fluxrec} may be used to recover the associated interface fluxes of the primary system upon first supplying the boundary fluxes\footnote{Recall that in most cases, Eq.~\eqref{eq:FDEC_bcflux} is just an evaluation of the flux at the boundary nodes and that only generalized SBP operators \cite{DRF:2014} feature specialized boundary extrapolation operators that would yield the split-form averaging of the flux at the boundary.}, which are the following for the proposed split form (here focusing on the left boundary for brevity):
\begin{eqnarray}
    \mathcal{F}_\ell^{(n)} = \left[\begin{array}{c} 
    \mathcal{F}_{\rho} \\ \\ \mathcal{F}_{\rho u_m} \\ \\ \mathcal{F}_{\rho E}
    \end{array} \right]_\ell^{(n)} =\left[
    \begin{array}{c}
    (\tilde{\rho} u_n)_\ell \\ \\
    \frac{1}{2}(\tilde{\rho} u_n u_m)_\ell + \frac{1}{2}(\tilde{\rho} u_n)_\ell (u_m)_\ell + \delta_{mn} \cdot p_\ell \\ \\
    \frac{1}{2} \sum_{m=1}^d (u_m)_\ell(\tilde{\rho} u_n u_m)_\ell + (\tilde{\rho} u_n \tilde{e})_\ell + (u_n)_\ell p_\ell
    \end{array}
    \right] .\label{eq:FDEC_bcflux}
\end{eqnarray} 
These boundary fluxes may further be used to employ necessary boundary and block interface coupling via Simultaneous Approximation Terms (SATs) \cite{Zingg:2014}.

It is insightful to also observe the structure of the resulting fluxes on the domain interior.
For example, the two-point central-difference stencils give the following ``wide-width" interior interface fluxes
\begin{eqnarray}
&&\left( \boldsymbol{\mathcal{C}}^{(n)} + \boldsymbol{\mathcal{P}}^{(n)}\right)_{\text{2pt}}  \nonumber \\ \nonumber \\
 && \to \ \mathcal{F}^{(n)}_{i+1/2}\big|_{\text{wide}} 
    = 
\left[
    \begin{array}{c}
    \frac{1}{2}\overline{(\rho_i,\rho_{i+2})}^{\log,n} u_{i+1} +     \frac{1}{2}\overline{(\rho_{i-1},\rho_{i+1})}^{\log,n} u_{i}  \\ \\
    \mathcal{F}_{\rho,i+1/2}^{(n)} \cdot \frac{(u_{m,i+1} + u_{m,i})}{2} + \delta_{mn} \cdot \frac{p_{i+1} + p_i}{2}  \\ \\
    \mathcal{F}_{\rho,i+1/2}^{(n)} \cdot\sum_{m=1}^d \frac{u_{m,i} u_{m,i+1}}{2} \\
    + \ \frac{1}{2}\overline{(\rho_i,\rho_{i+2})}^{\log,n} u_{i+1}\overline{(e_i,e_{i+2})}^{H\log,n} \\ + \ \frac{1}{2}\overline{(\rho_{i-1},\rho_{i+1})}^{\log,n} u_{i}\overline{(e_{i-1},e_{i+1})}^{H\log,n}  \\ + \ \frac{u_{n,i}p_{n,i+1} + u_{n,i+1}p_{n,i}}{2}  
    \end{array}
    \right] \label{eq:ECFlux_wide}
\end{eqnarray} 
and the two-point biased difference stencils yield the following parameterized ``narrow-width" interior interface fluxes:
\begin{eqnarray}
&&\omega \cdot \left( \boldsymbol{\mathcal{C}}^{(n,+)} + \boldsymbol{\mathcal{P}}^{(n,+)}\right)_{\text{2pt}} + (1-\omega) \cdot \left( \boldsymbol{\mathcal{C}}^{(n,-)} + \boldsymbol{\mathcal{P}}^{(n,-)}\right)_{\text{2pt}} \nonumber \\ \nonumber \\
 && \to \ \mathcal{F}^{(n)}_{i+1/2}\big|^\omega_{\text{narrow}} 
    =
\left[
    \begin{array}{c}
    \overline{(\rho_i,\rho_{i+1})}^{\log,n} \left((1-\omega) \cdot u_{n,i} + \omega \cdot u_{n,i+1} \right) \\ \\
    \mathcal{F}_{\rho,i+1/2}^{(n)} \cdot \frac{(u_{m,i+1} + u_{m,i})}{2} + \delta_{mn} \cdot \left((1-\omega) \cdot p_{i+1} + \omega  \cdot p_i\right) \\ \\
    \mathcal{F}_{\rho,i+1/2}^{(n)} \cdot \left(\sum_{m=1}^d \frac{u_{m,i} u_{m,i+1}}{2} + \overline{(e_i,e_{i+1})}^{H\log,n} \right) \\ + \ (1-\omega) \cdot u_{n,i}p_{i+1} + \omega \cdot u_{n,i+1}p_i
    \end{array}
    \right] . \label{eq:ECFlux_narrow}
\end{eqnarray} 
In the above, the arguments of the nonlinear averages computed in the $n$ direction are explicitly written for additional clarity.
Note that while the two-point wide scheme stems from two-point differences, the resulting flux is multi-point.

Some interesting insights transpire from studying the interior flux forms stemming from the two-point schemes. 
First, we note that the central stencils induce a wider node dependence and furthermore that their fluxes couple the thermodynamic interface reconstructions with that of the transport velocity.
On the other hand, the biased stencils---at least in the two-point scenario---split up these variable reconstructions.
Next, we note that the biased stencils yield counter-balanced biasing in the flux with respect to velocity and pressure when $\omega \ne 1/2$.
The associated interface reconstructions in such cases are of lower order than the symmetric case ($\omega = 1/2$), as expected by the fact that such biased stencils are typically one order lower than their centered counterparts.
To the authors' knowledge, this is the first time that EC fluxes with biased interior quantities have been presented, and leveraging such properties toward improving the scheme resolution of sharp gradients constitutes a line of future research.

Also notable upon studying these interior fluxes is the fact that a symmetric combination of the biased two-point operators with $\omega = 1/2$ recovers the second-order flux of Ranocha \cite{Ranocha:2018phd,Ranocha_CAMC_2021},
therefore formalizing its finite-difference compatibility. 
This exact congruence between the symmetric two-point narrow-width schemes and the Ranocha flux is not expected to hold to high-order, however.
Namely, the Ranocha flux (as well as other two-point EC fluxes in the literature~\cite{Ranocha_JSC_2018,Winters_BIT_2020}) is typically extended to high-order via flux-differencing~\cite{LeFloch:2002,Fisher_JCP_2013b}, which is a Richardson-type extrapolation of the low-order two-point fluxes.
This, however, need not be the case for the current FD-EC methods which can technically be extended to high-order simply by considering multi-point FD stencils in the definitions of $\tilde{\rho}$ and $\tilde{e}$ as well as the flux derivatives.
This is exemplified most simply when inspecting the mass flux term, which motivates the general uniqueness of the respective representations at high order.
Assuming a $K$-width central stencil with $p$-point biasing for the associated dual pair schemes, one would have
\begin{eqnarray}
     && \text{(via current FD-EC wide):} \nonumber \label{eq:FDECwide_massHO}\\
    &&\mathcal{F}_{\rho,i+1/2}^{(n)} - \mathcal{F}_{\rho,i-1/2}^{(n)} \nonumber \\
&& = \sum_{k=-K}^K d_{i,i+k}^{\text{c}} \cdot \left[\tilde{\rho}_{i+k}^{(n)}  \cdot u_{n,i+k}\right] \nonumber  \\
&& = \sum_{k=-K}^K d_{i,i+k}^{\text{c}} \cdot \left[\tilde{\rho}_{i}^{(n)}  \cdot u_{n,i} + \tilde{\rho}_{i+k}^{(n)}  \cdot u_{n,i+k} \right]  \\
&&= \sum_{k=-K}^K d_{i,i+k}^{\text{c}} \cdot \left[\overline{(\rho_{i+k-K},\dots,\rho_{i+k+K})}^{\log,n}   \cdot u_{n,i+k}  \right]    \nonumber  \\ \nonumber \\
   && \text{(via current FD-EC symmetric narrow, $\omega =1/2$):} \label{eq:FDECnarrow_massHO} \nonumber\\
    &&\mathcal{F}_{\rho,i+1/2}^{(n)} - \mathcal{F}_{\rho,i-1/2}^{(n)} \nonumber  \\
&& = \sum_{k=-K}^K  \frac{1}{2}\left(d_{i,i+k}^+ \cdot \tilde{\rho}_{i+k}^{(n,-)} + d_{i,i+k}^- \cdot \tilde{\rho}_{i+k}^{(n,+)} \right) \cdot u_{n,i+k} \nonumber \\ 
&& = \sum_{k=-K}^K  \frac{1}{2}\left(d_{i,i+k}^+ \cdot \left[\tilde{\rho}_{i}^{(n,-)} \cdot u_{n,i} + \tilde{\rho}_{i+k}^{(n,-)} \cdot u_{n,i+k} \right] + d_{i,i+k}^- \cdot \left[\tilde{\rho}_{i}^{(n,+)} \cdot u_{n,i} +  \tilde{\rho}_{i+k}^{(n,+)} \cdot u_{n,i+k}  \right]\right)  \\ \nonumber
&&= \sum_{k=-K}^K \frac{1}{2}\left(d_{i,i+k}^+ \cdot \overline{(\rho_{i+k-K},\dots,\rho_{i+k+K-p})}^{\log,n} + d_{i,i+k}^- \cdot \overline{(\rho_{i+k-K+p},\dots,\rho_{i+k+K})}^{\log,n}  \right) \cdot u_{n,i+k} \nonumber \\ \nonumber
    \\
&&\text{(via two-point flux-differencing):} \nonumber \label{eq:Ranocha_mass_fluxdiff}\\
    &&\mathcal{F}_{\rho,i+1/2}^{(n)} - \mathcal{F}_{\rho,i-1/2}^{(n)} \nonumber \\
    &&= \sum_{k=-K}^K d_{i,i+k}^{\text{c}} \cdot \left[ \overline{(\rho_i,\rho_{i+k})}^{\log,n} \cdot (u_{n,i} + u_{n,i+k})\right]
\end{eqnarray}
where in the above we have the following relations for the stencil coefficients
\begin{equation}
\sum_k d_{i,i+k}^{\text{c}} = \sum_k d_{i,i+k}^\pm =0 \quad \text{and} \quad \frac{(d^+_{i,i+k} + d^-_{i,i+k})}{2} = d_{i,i+k}^{\text{c}} 
\end{equation}
and where 
\begin{equation}
\text{(on the interior)}: \quad d_{i,i-k}^{\text{c}} = - d_{i,i+k}^{\text{c}} \ \text{and} \ d_{i,i}^{\text{c}} = 0 \ .
\end{equation}
Note that Eqs.~\eqref{eq:FDECwide_massHO} and \eqref{eq:FDECnarrow_massHO} are still expressed in a flux-difference flavor (albeit with multi-point information), which stems naturally from their finite difference origins.
Due to the nonlinearity of the logarithmic mean, it is clear that the standard flux-differenced implementation of Ranocha's method will differ from the newly presented symmetric narrow-width method for $K \ne 1$, as the latter would utilize a multi-point logarithmic filters of density rather than two-point logarithmic means.
Suitable algorithms are still under development for handling such multi-point nonlinear averages near singular points (i.e., where the quotient formulas from Section~\ref{sec:2} are near zero) and to properly enforce positivity of the output; therefore, the numerical results in the following section will focus on the class of two-point FD-EC methods.
Appendix \ref{sec:app_d}, however, provides an alternate error-reducing treatment  for the narrow biased schemes based on flux differencing with multi-point difference stencils.

%-----
\subsection{On the kinetic-energy and pressure-equilibrium preservation of the new schemes} \label{sec:4p1}

Besides being entropy conserving, the new methods feature additional robustness-promoting characteristics---namely, kinetic-energy and pressure-equilibrium preservation. 
These are briefly described here.

The FD-EC methods employ splittings for the mass and momentum equations that yield the kinetic energy preserving (KEP) formulation of Feiereisen \cite{Feiereisen_1981}, wherein no spurious convective dynamics are induced in the associated discrete secondary equation.
For example, the centered representation gives 
\begin{eqnarray}
     \boldsymbol{\mathcal{C}}_{\rho k} &=& \sum_{m=1}^d\left[u_m\boldsymbol{\mathcal{C}}_{\rho u_m} - \frac{1}{2} u_m^2\boldsymbol{\mathcal{C}}_\rho\right] \nonumber\\
\text{(centered)}     &=& \frac{1}{2} \left[u_m\delta_{x_n} (\tilde{\rho} u_n u_m) + \tilde{\rho} u_n u_m \delta_{x_n} u_m  \right] \\
&& \xrightarrow[]{\text{(2pt)}} \mathcal{F}^{(n)}_{\rho k,i+1/2} = \frac{1}{4} \left[(\tilde{\rho} u_n u_m)_i (u_m)_{i+1} + (\tilde{\rho} u_n u_m)_{i+1} (u_m)_{i}  \right] \nonumber
\end{eqnarray}
We note that the convective term in the kinetic energy equation is in a quadratically split form  which will result in local conservation of the quantities. 
Analogous expressions can be written for the biased stencils.
Although kinetic energy preservation does not provide a rigorous nonlinear stability metric for compressible flows, incorporating its numerical consistency has been shown to greatly enhance solution quality---even for the class of entropy-conserving Tadmor fluxes~\cite{Ranocha_JSC_2018,Chandrashekar_CCP_2013}.
And as motivated in Eqs.~\eqref{eq:EnergyConst_C} and \eqref{eq:EntVar_Euler_contraction} of Section~\ref{sec:2}, it forms a key consistency within the continuous entropy analysis.

Next, it is possible to also show that all the new FD-EC schemes have the Pressure Equilibrium Preservation (PEP) property~\cite{Shima_JCP_2021,Ranocha_CAMC_2021,DeMichele_JCP_2024} for the current context of a calorically perfect single fluid.
In other words, they have the ability to preserve the equilibrium of velocity and pressure across an interface.
The condition on the discretization of the momentum equation assuming constant velocity $U_o$ and pressure $P_o$ (see~\cite{DeMichele_JCP_2024, Ranocha_CAMC_2021}) is that the momentum flux is equal to
\begin{equation}\label{eq:PEP_condition_rhou_convective_flux}
    \mathcal{F}_{\rho u_m,i+1/2} = \mathcal{F}_{\rho,i+1/2} \cdot U_o + \mathop{\text{const($U_o, P_o$})}.
\end{equation}
This is clearly satisfied by the schemes considered---for example, see Eqs.~\eqref{eq:ECFlux_wide} and \eqref{eq:ECFlux_narrow}.
The other condition, on the internal-energy flux, is that it should be a constant dependent only on the values of $U_o$ and $P_o$ \cite{DeMichele_JCP_2024}: 
\begin{equation}
    \mathcal{F}_{\rho e,i+1/2} = \text{const($U_o$,$P_o$)}.
\end{equation}
This is clearly the case for the pressure-velocity term in internal energy ($p \delta_{x_n} u$) and it is also true for the convective flux since
\begin{equation}\label{eq:Flux_eint_upcost}
    \mathcal{F}^{(n,\text{conv})}_{\rho e,i+1/2} = \logmeanp{\rho}\,(\logmeanp{e^{-1}})^{-1} U_o =\logmeanp{\rho}\,(\logmeanp{\rho})^{-1} \frac{P_o U_o}{\gamma -1} \ .
\end{equation}
The ability to appropriately preserve such equilibrium conditions across interfaces is essential towards avoiding the numerical instigation of solution instabilities that can threaten both numerical robustness and solution quality~\cite{Fujiwara_JCP_2023,Ching:2024}.

\subsection{On the damping effects of split-form stencil biasing}
The schemes presented in the current section represent an array of different biasing and stencil dependencies, which are then expected to impact the resulting solution.
Qualitatively understanding the impact of such choices is possible upon parameterizing the stencil biasing as a central component plus an artificial dissipation term,
\begin{equation}
    \delta^{(\omega)} \triangleq \delta^c + \omega \cdot \delta^{AD}, \ \text{with} \ \omega \in [-1,1] \ \to \ \delta^{(\omega = 0)} = \delta^c, \ \delta^{(\omega = \pm 1)} = \delta^\pm \ . \label{eq:fd_bias}
\end{equation}
Note that the above is simply a rearrangement of Eq.~\eqref{eq:cent_dp_rel}, with $\omega$ controlling the level of biasing.
Such a re-writing, however, allows one to highlight the anticipated damping effects inherent within the biased split-forms, as will be shown next.

The overall group of proposed finite-difference compatible entropy-conserving (FD-EC) methods can be expressed in semi-discrete residual form as
\begin{equation}
    d_t q = \alpha \cdot \mathcal{R}^{(\omega)} + (1 - \alpha) \cdot \mathcal{R}^{(-\omega)} , \ \text{where} \ \mathcal{R}^{(\omega)} = -(\mathcal{C}^{(\omega)} + \mathcal{P}^{(\omega)}) \ , \label{eq:FD-EC_2param}
\end{equation}
with the convective and pressure terms summarized in Table~\ref{table:3} and depending on the biasing parameter $\omega$.
Equation~\eqref{eq:FD-EC_2param} is thus a two-parameter family of schemes based on the pair $[\alpha,\omega]$.
The centered wide stencils are recovered for the parameter pair $(\alpha, \omega = 0)$; the centered narrow stencils are recovered for $(\alpha = 1/2, \omega \ne 0)$; and, the ‘‘fully" biased stencils (i.e., at least relative to velocity and pressure) are recovered for $(\alpha=\pm 1, \omega = \pm 1)$.

\begin{table}[h!]
\center 
\begin{tabular}{l | l }
  &  \\ \hline   \\
$\boldsymbol{\mathcal{C}}_{\rho}^{(n,\omega)} $ & $ \delta^{(\omega)}_{x_n} (\tilde{\rho}^{(-\omega)} u_n)$ \\ \\
$\boldsymbol{\mathcal{C}}_{\rho u_m}^{(n,\omega)}$& $ \frac{1}{2}\left[\delta_{x_n}^{(\omega)} (\tilde{\rho}^{(-\omega)} u_n u_m) + \tilde{\rho}^{(-\omega)} u_n \delta_{x_n}^{(-\omega)} u_m + u_m \delta_{x_n}^{(\omega)} (\tilde{\rho}^{(-\omega)} u_n) \right]$ \\ \\
$\boldsymbol{\mathcal{C}}_{\rho e}^{(n,\omega)}$ & $ \delta^{(\omega)}_{x_m} (\tilde{\rho}^{(-\omega)} u_n \tilde{e}^{(-\omega)})$  \\ \\
$\boldsymbol{\mathcal{P}}_{\rho u_m}^{(n,\omega)}$ & $  \delta^{(-\omega)}_{x_m} p$ if $n=m$, otherwise $0$  \\ \\
$\boldsymbol{\mathcal{P}}_{\rho e}^{(n,\omega)}$ & $ p\delta^{(\omega)}_{x_n} u_n$ \\ \\
\end{tabular}
\caption{A kinetic-energy and pressure-equilibrium preserving finite-difference discretization of Euler derivatives assuming SBP-compatible stencils. The auxiliary variables are 
$\tilde{\rho}^{(\omega)} \triangleq \frac{\delta_{x_n}^{(\omega)} \rho}{\delta_{x_n}^{(\omega)} \log \rho} \equiv \overline{\rho}^{(\log,n,\omega)}$ and 
$\tilde{e}^{(\omega)} \triangleq \frac{\delta^{(\omega)} \log e^{-1}}{\delta^{(\omega)} e^{-1}} \equiv \overline{e}^{(H\log,n,\omega)}$. Centered wide stencil method is recovered for $\omega = 0$.} \label{table:3}
\end{table}

Leveraging Eq.~\eqref{eq:fd_bias}, the split-form biasing associated with the discretized residual can be decomposed into ``central" and ``dissipation" components for a given bias parameter $\omega$, $\mathcal{R}^{(\omega)} = (\mathcal{R}^{(\omega),c} + \mathcal{R}^{(\omega),AD})$:
\begin{align}
    -\mathcal{R}_n^{(\omega),c} &= 
    \left[\begin{array}{l}
    \delta_{x_n}^c \tilde{\rho}^{(-\omega)} u_n  \\ \\
    \frac{1}{2} \left[\delta_{x_n}^c \tilde{\rho}^{(-\omega)} u_nu_m + \tilde{\rho}^{(-\omega)} u_n \delta_{x_n}^c u_m + u_m \delta_{x_n}^c \tilde{\rho}^{(-\omega)} u_n\right] + \delta_{x_n}^c p 
    \\ \\
    \delta_{x_n}^c \tilde{\rho}^{(-\omega)} u_n\tilde{e}^{(-\omega)} + \frac{1}{2} \left[u_m \delta_{x_n}^c \tilde{\rho}^{(-\omega)} u_n u_m + \tilde{\rho}^{(-\omega)} u_n u_m \delta_{x_n}^c u_m \right]  + (u_m \delta_{x_n}^c p + p \delta_{x_n}^c u_m) 
    \end{array}\right] \label{eq:FD-EC_Res_c}\\ \nonumber \\
    -\mathcal{R}_n^{(\omega),AD} &=  \left[\begin{array}{l}
    \omega \cdot \delta_{x_n}^{AD} \tilde{\rho}^{(-\omega)} u_n \\ \\
     \omega \cdot \left[\frac{1}{2}\delta_{x_n}^{AD} \tilde{\rho}^{(-\omega)} u_nu_m +  \frac{1}{2}\left(u_m \delta_{x_n}^{AD} \tilde{\rho}^{(-\omega)} u_n - \tilde{\rho}^{(-\omega)} u_n \delta_{x_n}^{AD} u_m\right) - \delta_{x_n}^{AD} p   \right] \\ \\ 
     \omega \cdot \left[ \delta_{x_n}^{AD}  \tilde{\rho}^{(-\omega)} u_n\tilde{e}^{(-\omega)} + (\tilde{\rho}^{(-\omega)} u_n u_m \delta_{x_n}^{AD} u_m - u_m \delta_{x_n}^{AD} \tilde{\rho}^{(-\omega)} u_n u_m) + (p \delta_{x_n}^{AD} u_m - u_m \delta_{x_n}^{AD} p) \right] \ .
    \end{array}\right] \label{eq:FD-EC_Res_AD}
\end{align}
From the above, one can anticipate the narrow split forms to produce an additional damping effect in the density and pressure fields compared to the wide-stencil configuration of $\omega = 0$. 
This will occur even though the overall formulation is kinetic-energy-preserving and entropy-conserving, and it may be more noticeable in flows featuring sharp gradients.
The narrow symmetric implementation (i.e., $(\alpha = 1/2,\omega \ne 0)$ will lack additional pressure damping compared to the biased renditions.
Namely, the narrow symmetric scheme lacks $\delta_{x_n}^{AD}p$ in the momentum equation as well as $(p \delta_{x_n}^{AD} u_m - u_m \delta_{x_n}^{AD} p)$  
in the total energy equation.
Such nuances highlight the distinctness of the respective schemes, which is further explored in the following numerical evaluations.

%-----
\section{Numerical results} \label{sec:5}

The new EC schemes are here tested with a variety of cases (i.e., the 2D isentropic vortex, the 1D density wave, the 1D Sod shock tube, and the 3D Taylor--Green vortex) in order to verify the validity of the theoretical results.
In particular, we consider the new scheme based on the wide stencil presented in Eq.~\eqref{eq:ECFlux_wide}, here referred to as \ECw, and three schemes from the family identified in Eq.~\eqref{eq:ECFlux_narrow}, that with $\omega=0$, $\omega=1$, and $\omega =0.5$ are labeled \ECb, \ECf, and \ECs,  respectively. 
Notably, \ECs is analogous to the scheme of Ranocha in the scenario of a two-point second order stencil.
A summary of the schemes and their respective numerical fluxes is presented in Table~\ref{tab:schemes}.

Despite having demonstrated the capability to obtain high-order formulations of the presented EC schemes, the methods will all be tested with their low-order version, employing first and second order accurate differential operators.
This is due to the singularity problem that occurs when evaluating the logarithmic mean for almost identical values.
This challenge is handled in the case of two-point means by employing suitable expansion approximations such as the one famously introduced by \citet{Ismail_JCP_2009}. 
Other similar approaches have been recently employed to produce schemes that are asymptotically entropy conserving \cite{Tamaki_JCP_2022, DeMichele_JCP_2023,Kawai:2025}.
However, a generalization of the procedure for multiple-point filters has yet to be developed.

The results presented in this section have been generated based on the flux forms of the FD schemes in the code STREAmS-2~\cite{Bernardini_CPC_2023}.
Density, momentum and total energy are employed as the primary variables. 
All time integrations have been performed using the classical explicit fourth-order Runge-Kutta (RK4) scheme.
In order to analyze the conservation properties of the schemes, we introduce the  notation of using the angular brackets to indicate integration over the spatial domain $\langle\phi\rangle = \int_{\Omega} \phi \, \d\Omega$ and using $\epsilon$ for the nondimensionalization $\epsilon_\phi = \left(\langle{\phi}\rangle - \langle{\phi_0}\rangle\right)\left/\left|\langle{\phi_0}\rangle\right|\right.$, in which $\phi_0$ indicates the values at the starting time.
For all the simulations, the value for the normalized specific heat capacity at constant volume is  $c_v=1$ and $\gamma =1.4$.
 \begin{table}
\renewcommand\arraystretch{1.8}
\centering
\begin{adjustbox}{max width=\textwidth}
\begin{tabular}{cccccc}
&Ref.&&$\mathcal{F}_{\rho}^{(n)}$ &   $\mathcal{F}_{\rho u_m,i+1/2}^{(n)}$ &   $\mathcal{F}_{\rho E}^{(n)}$  \\
\hline
\ECb & new &&$ \logmeanp{(\rho_i,\rho_{i+1})} u_{n,i}$& $\mathcal{F}_{\rho,i+1/2} \meanp{u_m} + p_{i+1}$& $\mathcal{F}_{\rho}\hlogmeanp{(e_i,e_{i+1})} + \mathcal{F}_{\rho}\,\sum_m \frac{u_{m,i} u_{m,i+1}}{2} + p_{i+1}u_{n,i}$ \\
\ECf & new &&$ \logmeanp{(\rho_i,\rho_{i+1})} u_{n,i+1}$& $\mathcal{F}_\rho \meanp{u_m} + p_{i}$& $\mathcal{F}_\rho\hlogmeanp{(e_i,e_{i+1})} + \mathcal{F}_\rho\,\sum_m \frac{u_{m,i} u_{m,i+1}}{2} + p_{i}u_{n,i+1}$ \\
\ECs & \cite{Ranocha_JSC_2018} &&$ \logmeanp{(\rho_i,\rho_{i+1})} \meanp{u_{n}}$& $\mathcal{F}_\rho \meanp{u_m} + \meanp{p}$& $\mathcal{F}_\rho\hlogmeanp{(e_i,e_{i+1})} + \mathcal{F}_\rho\,\sum_m \frac{u_{m,i} u_{m,i+1}}{2} + \pmean{(p,u)}$ \\
\multirow{3}{*}{\ECw} & \multirow{3}{*}{new} &&$ 0.5 \logmeanp{(\rho_{i-1},\rho_{i+1})} u_{n,i}$& \multirow{3}{*}{$\mathcal{F}_\rho \meanp{u_m} + \meanp{p}$}& $0.5\logmeanp{(\rho_{i-1},\rho_{i+1})} u_{n,i}\hlogmeanp{(e_{i-1},e_{i+1})} $ \\
&&& $+0.5 \logmeanp{(\rho_{i},\rho_{i+2})} u_{n,i+1}$&&$+0.5\logmeanp{(\rho_{i},\rho_{i+2})} u_{n,i+1}\hlogmeanp{(e_i,e_{i+2})} $\\
&&&&&$+ \mathcal{F}_\rho\,\sum_m \frac{u_{m,i} u_{m,i+1}}{2} + \pmean{(p,u)}$\\
\hline
\end{tabular}
\end{adjustbox}
\caption{
Numerical fluxes for the schemes used for the tests. The fluxes are evaluated along direction $n$ at location $(i+1/2)$. All interpolation operations are implicitly interpreted to be in the same direction $n$.
} \label{tab:schemes} 
\end{table}

%------------
\subsection{2D isentropic vortex}
The two-dimensional isentropic Euler vortex problem has initial conditions
\begin{align}
    \frac{ u(x,y) }{u_\infty} &= 1 - \frac{M_v}{M_\infty}\frac{y-y_0}{r_v}e^{(1-\hat{r}^2)/2} \nonumber \\
    \frac{ v(x,y) }{u_\infty} &= \frac{M_v}{M_\infty}\frac{x-x_0}{r_v}e^{(1-\hat{r}^2)/2} \nonumber \\
    \frac{ T(x,y) }{T_\infty} &=\left(\frac{ p(x,y) }{p_\infty}\right)^{(\gamma-1)/\gamma}  =\left(\frac{ \rho(x,y) }{\rho_\infty} \right)^{\gamma-1} = 1 - \frac{\gamma-1}{2} M_v^2 e^{1-\hat{r}^2} \nonumber
\end{align}
in which $\hat{r} = r/r_v$ with core radius $r_v = 1/15$ and $r$ is the distance from the vortex center.
The characteristic values of velocity, density, and temperature are $u_\infty =1,\ \rho_\infty=1,\   T_\infty = 720 M_v r_v$, while pressure is derived from density as $p = \rho^\gamma / (\gamma M_\infty ^2)$.
The mean flow Mach number is $M_\infty = 0.5$ and 
the vortex, whose center is initially at coordinates $(x_0,y_0) = (0.5,0.5)$, has a strength $M_v = 0.5$.
The vortex core is $\rho(x_0,y_0) = 0.6941 \rho_\infty$; a non-negligible perturbation that is meant to instigate nonlinear effects. 
A $30\times 30$ discretization is used for the square domain of unitary side with a uniform Cartesian grid; boundary conditions are periodic in all directions.
Upon choosing to identify the vortex width as the point where $\rho/\rho_\infty = 0.99$, then approximately eight vortex widths can be said to fit along the length of the domain.
The Courant number of the tests is set to $\text{CFL} = 0.001$, corresponding to a time step size $\Delta t=2\times10^{-5}$.
A full loop of the domain (i.e., approximately eight vortex widths) occurs each $t = 1$.

\begin{figure}[tb]
     \centering
    \begin{subfigure}[b]{0.47\textwidth}
         \centering
         \includegraphics[width=\textwidth]{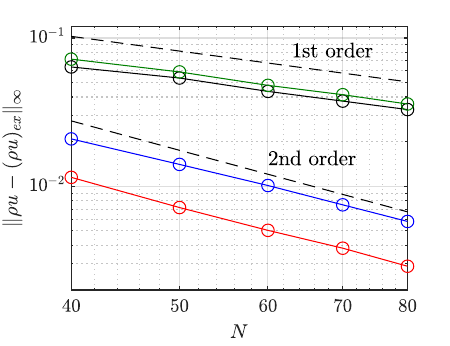}
         \caption{Order of accuracy on solution of $\rho u$}
         \label{fig:IV_accuracy}
     \end{subfigure}
    \begin{subfigure}[b]{0.47\textwidth}
         \centering
         \includegraphics[width=\textwidth]{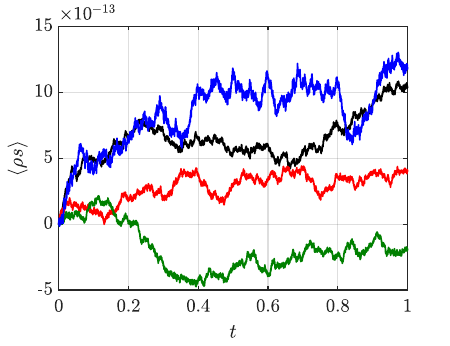}
         \caption{Time evolution of entropy integral}
         \label{fig:IV_entropy}
     \end{subfigure}
    \caption{Isentropic vortex test for different discretizations: black lines are used for \ECb, red for \ECs, green for \ECf, blue \ECw. $\text{CFL} = 0.01$. }
    \label{fig:IV}
\end{figure}
The isentropic vortex flow is an exact solution of the inviscid compressible flow equations and as such it can be used to evaluate the accuracy of the numerical methods.
In Fig.~\ref{fig:IV_accuracy} the error on momentum along the $x$ axis is displayed as a function of $N$ number of nodes in each direction.
The error is evaluated as the maximum norm of the difference between the computed and the exact solution for momentum along the $x$ direction $\|\rho u- (\rho u)_{ex}\|_\infty$, evaluated at a time $t=0.01$ with $\CFL=0.01$. 
The analysis shows a difference between the first-order accurate (\ECf and \ECb) and the second-order accurate (\ECs and \ECw) methods.
The reason behind this behavior is easily explained by the fact that \ECf and \ECb use an asymmetrical numerical derivative for the discretization of the pressure term and an asymmetrical interpolator for velocity as well; on the other hand both \ECs and \ECw result in central schemes.
Identical trends are witnessed upon inspecting the error of other quantities (not shown), such as the primitive variables $(\rho,u,T)$.

Despite this difference on the order of accuracy, all the schemes are able to exactly conserve entropy in a discrete sense, with an error in the conservation of the order of machine zero.
The methods present very similar behavior, with a slightly larger drift observed in the asymmetric schemes.

%---
\subsection{1D density wave}

\begin{figure}[h]
     \centering
    \begin{subfigure}[b]{0.47\textwidth}
         \centering
         \includegraphics[width=\textwidth]{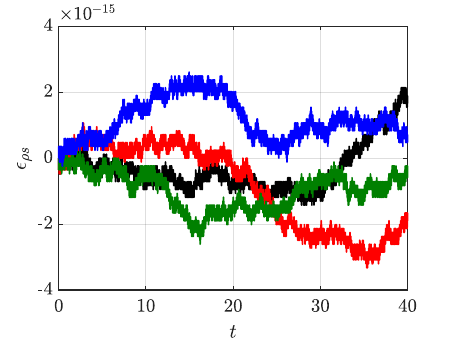}
         \caption{Entropy evolution}
         \label{fig:DW_rhos}
     \end{subfigure}
    \begin{subfigure}[b]{0.47\textwidth}
         \centering
         \includegraphics[width=\textwidth]{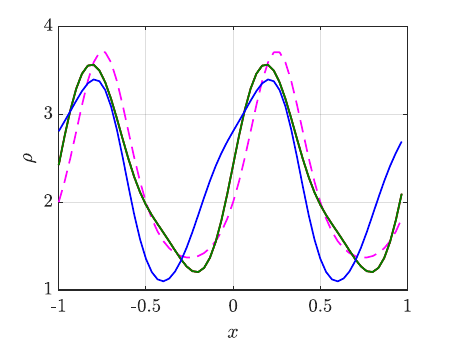}
         \caption{Density}
         \label{fig:DW_rho}
     \end{subfigure}
    \begin{subfigure}[b]{0.47\textwidth}
         \centering
         \includegraphics[width=\textwidth]{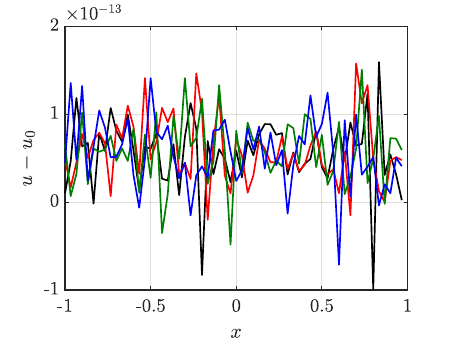}
         \caption{Velocity}
         \label{fig:DW_u}
     \end{subfigure}
    \begin{subfigure}[b]{0.47\textwidth}
         \centering
         \includegraphics[width=\textwidth]{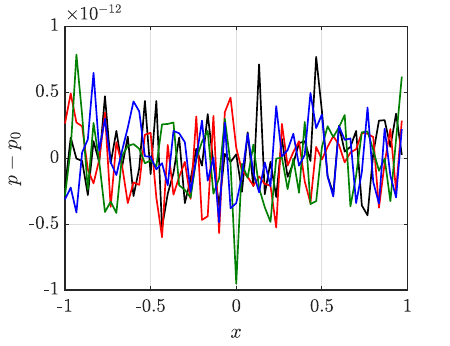}
         \caption{Pressure}
         \label{fig:DW_p}
     \end{subfigure}
    \caption{Density wave test for different discretizations: black lines are used for \ECb, red for \ECs, green for \ECf, blue \ECw; dashed magenta line shows the exact solution for density. $\text{CFL} = 0.01$, final solutions at $T=40$.}
    \label{fig:DW}
\end{figure}

It can be shown theoretically (see Section~\ref{sec:4p1}) that all of the current schemes have the PEP property.
The simulation of a travelling density wave is well suited to numerically test the PEP property of the schemes.
The initial conditions for the test are
\begin{equation*}
    \rho_0 = 1 + \exp\left(\sin\left(\frac{2\pi x}{L}\right)\right), \qquad u_0 = 1, \qquad p_0 = 1,
\end{equation*}
with the domain $[-L, L]$ with $L = 1$ discretized in $61$ points and periodic boundary conditions. 
The Courant number has been chosen to be $\CFL = 0.001$ to prevent errors on entropy conservation due to the temporal integrator, and the simulation has been carried out until a final time $T=40$, which corresponds to approximately 20 cycles of the wave through the domain.

The simulation of the density wave also confirms the schemes' ability to exactly conserve global entropy  (see Fig.~\ref{fig:DW_rhos}). 
The theoretical prediction that the schemes would be PEP has also been confirmed: the errors introduced on velocity $u$ and pressure $p$ remain at machine precision even at the final time, as shown in Figs.~\ref{fig:DW_u}--\subref{fig:DW_p}.
For this test, the results produced by the two-point schemes (\ECb, \ECf, and \ECs) are indistinguishable, as shown for the density in Fig.~\ref{fig:DW_rho}.
This is due to the fact that the only difference between these methods is in the interpolation of $u$ and $p$, which becomes irrelevant when they are constant throughout the domain.
On the other hand, the results obtained with the scheme \ECw are appreciably different, which is likely a consequence of the wider stencil.
%
%-------
\subsection{1D Sod shock tube}
\label{sec:Sod}

While the EC schemes have been derived in the hypothesis of smooth flows, it is important to observe their behavior in the presence of discontinuities, as this pushes the limits of the non-dissipative methods.
When applied in shocked regions, these schemes will still conserve entropy, although producing non-physical oscillations.

In this vein, the Sod shock tube test is considered.
The initial conditions upstream (U) and downstream (D) of the diaphragm, which is located at midpoint of the domain, are given by
\begin{align*}
    \rho_U &= 1;  &   u_U &= 0; &    p_U &= 1; \\
    \rho_D &= 0.125;  &   u_D &= 0;  &   p_D &= 0.1. 
\end{align*}
The simulation is carried out until $T=0.1$ with $\CFL = 0.01$ based on velocity $u_\infty = \sqrt{p_U/\rho_U}$.
For the discretization of the interval $[-0.25,0.25]$, $101$ nodes are employed, however a much larger domain has been used to prevent the periodic boundary conditions to have any effect on the solution.

\begin{figure}[hp!]
     \centering
    \begin{subfigure}[b]{0.47\textwidth}
         \centering
         \includegraphics[width=\textwidth]{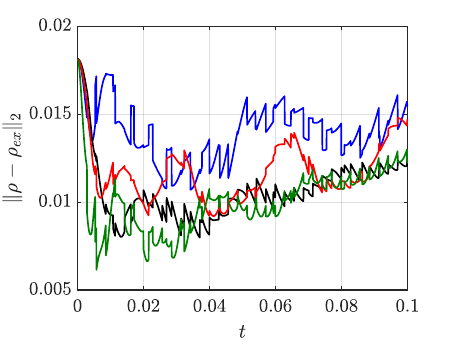}
         \caption{Density}
         \label{fig:Sod_rho_norm}
     \end{subfigure}
    \begin{subfigure}[b]{0.47\textwidth}
         \centering
         \includegraphics[width=\textwidth]{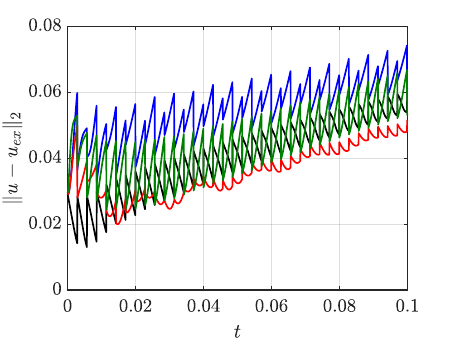}
         \caption{Velocity}
         \label{fig:Sod_u_norm}
     \end{subfigure}
    \begin{subfigure}[b]{0.47\textwidth}
         \centering
         \includegraphics[width=\textwidth]{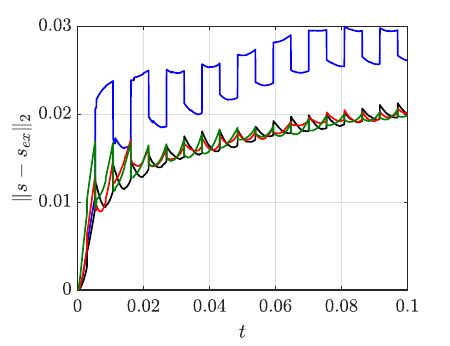}
         \caption{Entropy}
         \label{fig:Sod_s_norm}
     \end{subfigure}
    \begin{subfigure}[b]{0.47\textwidth}
         \centering
         \includegraphics[width=\textwidth]{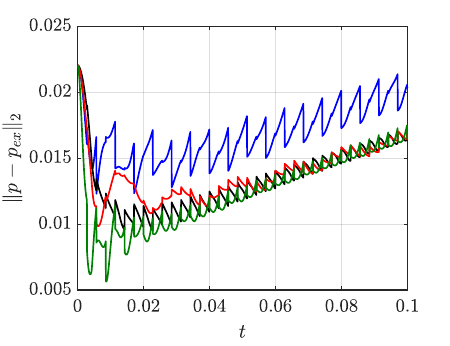}
         \caption{Pressure}
         \label{fig:Sod_T_norm}
     \end{subfigure}
    \caption{Sod test for different discretizations: black lines are used for \ECb, red for \ECs, green for \ECf, blue \ECw. $\CFL = 0.01$. Evolution in time of the $L_2$ norm of the difference between computed solution and analytical one. 
    }
    \label{fig:Sod_norm}
\end{figure}

In order to appreciate the difference between the various schemes,
Fig.~\ref{fig:Sod_norm} shows the evolution in time of the $L_2$ norm of the difference between the computed solution and the analytical one in the interval $[-0.25,0.25]$. The analytical solutions has been obtained with an exact Riemann solver.
While the \ECs scheme seems to attain better results considering velocity (Fig.~\ref{fig:Sod_u_norm}), this is no longer the case when other variables are examined.
In fact, at least one of the biased schemes, \ECb and \ECf, appears to be outperforming the symmetric schemes; this is especially evident in the case of density (Fig.~\ref{fig:Sod_rho_norm}).

\begin{figure}[hp!]
     \centering
    \begin{subfigure}[b]{0.47\textwidth}
         \centering
         \includegraphics[width=\textwidth]{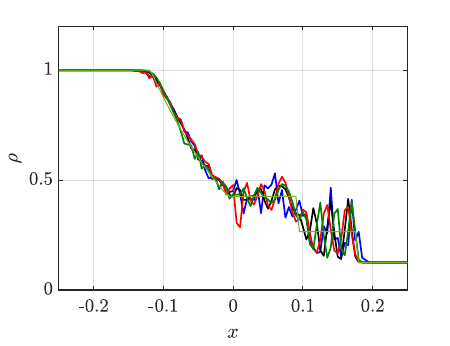}
         \caption{Density}
         \label{fig:Sod_rho}
     \end{subfigure}
    \begin{subfigure}[b]{0.47\textwidth}
         \centering
         \includegraphics[width=\textwidth]{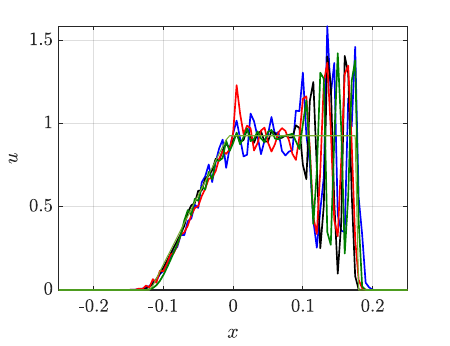}
         \caption{Velocity}
         \label{fig:Sod_u}
     \end{subfigure}
    \begin{subfigure}[b]{0.47\textwidth}
         \centering
         \includegraphics[width=\textwidth]{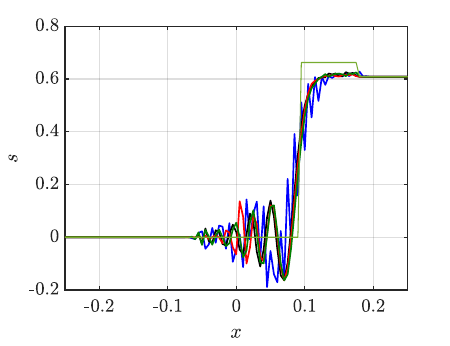}
         \caption{Entropy}
         \label{fig:Sod_s}
     \end{subfigure}
    \begin{subfigure}[b]{0.47\textwidth}
         \centering
         \includegraphics[width=\textwidth]{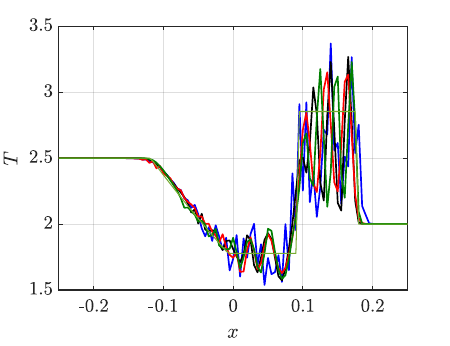}
         \caption{Temperature}
         \label{fig:Sod_T}
     \end{subfigure}
    \begin{subfigure}[b]{0.47\textwidth}
         \centering
         \includegraphics[width=\textwidth]{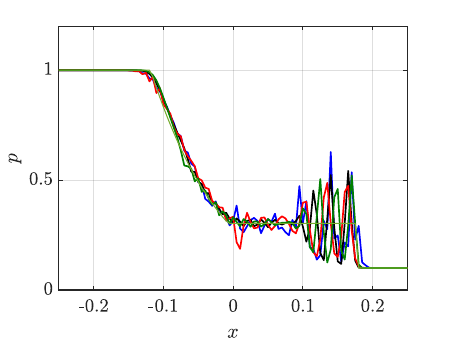}
         \caption{Pressure}
         \label{fig:Sod_p}
     \end{subfigure}
    \begin{subfigure}[b]{0.47\textwidth}
         \centering
         \includegraphics[width=\textwidth]{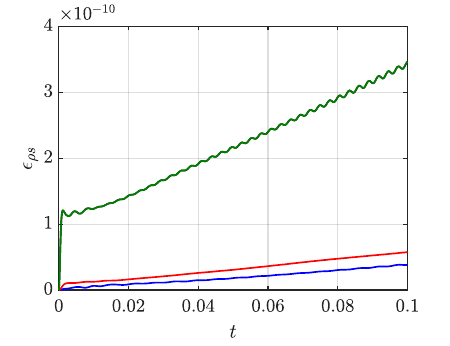}
         \caption{Entropy integral evolution}
         \label{fig:Sod_entropy}
     \end{subfigure}
    \caption{Sod test for different discretizations: black lines are used for \ECb, red for \ECs, green for \ECf, blue \ECw; pale green line is the exact solution. $\CFL = 0.01$, final solutions at $T=0.1$.}
    \label{fig:Sod}
\end{figure}

To evaluate the spatial distribution of the error, in Fig.~\ref{fig:Sod}, the solutions at time $T=0.1$ obtained with the various EC schemes are compared with the analytical one.
As expected, all the schemes are entropy conserving (Fig.~\ref{fig:Sod_entropy}) and generate spurious oscillations in the solution. 
As visible in Fig.~\ref{fig:Sod_s}, none of the schemes are able to reproduce the correct behavior for entropy in the region leading up to shock $x \in [0.1,0.2]$, and \ECw seems to present more oscillations. 
This behavior has been observed before for the scheme of Ranocha~\cite{Ranocha_CAMC_2021} (here \ECs) in~\cite{Edoh2024}.
This region is also one in which the oscillations are stronger for all the schemes when other variables are considered (e.g., density, velocity, temperature, pressure).

On the other hand, in the region $x\in[0,0.1]$ the asymmetric schemes \ECb and \ECf present oscillations with smaller amplitude when compared to the central schemes \ECw and \ECs. 
This is particularly evident when inspecting velocity in Fig.~\ref{fig:Sod_u}, for which a larger spike is present near $x=0$ for the \ECs scheme.
Such behavior is supported by the presence of additional damping terms in the density and pressure fields as suggested by Eqs.~\eqref{eq:FD-EC_Res_c} and \eqref{eq:FD-EC_Res_AD}. 
Despite being kinetic-energy-preserving and entropy-conserving, the split-form biasing inherently provides a low level of regularization compared to the centralized configurations.
This has also been observed in previous work for biased split-forms~\cite{Coppola_JCP_2023} and further motivates interest in studying these schemes for use in flows exhibiting sharp gradients.

While the focus of this work is on EC schemes, Appendix \ref{sec:app_c} demonstrates how they can be the base to Entropy Stable (ES) schemes through the addition of a specialized dissipative term. These ES schemes are tested again on Sod's test, to evaluate their ability to suppress oscillations in the presence of strong gradients.

%---
\subsection{3D Taylor-Green vortex}

The inviscid Taylor--Green vortex is used to test the schemes in a three-dimensional setting that serves as a benchmark for turbulence simulations; it generates smaller and smaller scales through vortex stretching.
The initial conditions are given by
\begin{align*}\label{eq:TGV_IC}
    \rho(x,y,z) &= \rho_0\nonumber\\
    u(x,y,z) &= u_0\sin(x)\cos(y)\cos(z)\nonumber\\
    v(x,y,z) &= -u_0\cos(x)\sin(y)\cos(z)\\
    w(x,y,z) &= 0\nonumber\\
    p(x,y,z) &= p_0 + u_0^2\frac{(\cos(2x) + \cos(2y))(2+\cos(2z))}{16}\nonumber
\end{align*}
in which $\rho_0=1$, $p_0=100$, $u_0=M_0 \sqrt{\gamma p_0/\rho_0}$.
Two different Mach numbers have been considered: for an almost incompressible case, $u_0=1$ which corresponds to $M_0\approx0.08$; for the case in which the effects of compressibility are taken into account, $M_0 = 0.4$ has been chosen.
The domain is a cube of length $2\pi$ with periodic boundary conditions in each direction.

%-------------

\subsubsection{Nearly incompressible case\texorpdfstring{ ($M_0\approx0.08$)}{}}

\begin{figure}[ph!]
     \centering
    \begin{subfigure}[b]{\textwidth}
         \centering
         \includegraphics[width=0.47\textwidth]{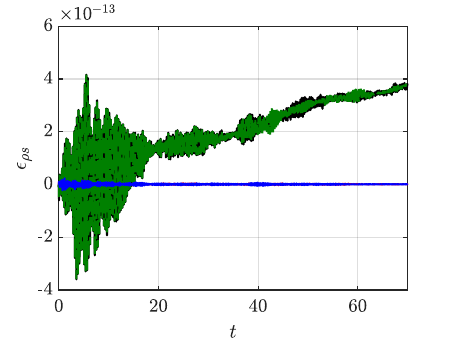}
         \includegraphics[width=0.47\textwidth]{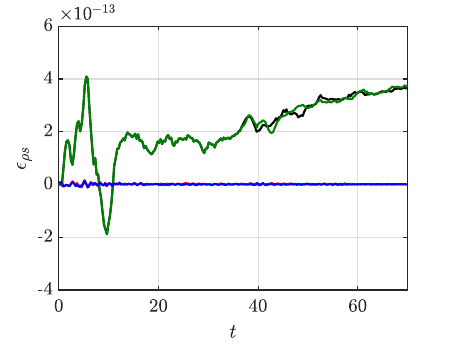}
         \caption{Time evolution of entropy integral}
         \label{fig:TGV_entropy}
     \end{subfigure}
    \begin{subfigure}[b]{\textwidth}
         \centering
         \includegraphics[width=0.47\textwidth]{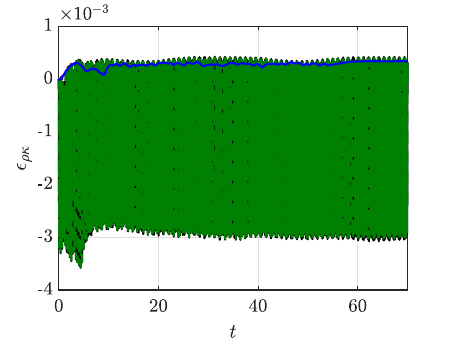}
         \includegraphics[width=0.47\textwidth]{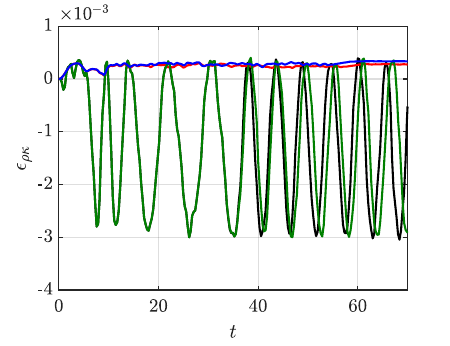}
         \caption{Time evolution of kinetic energy integral}
         \label{fig:TGV_kinetic}
     \end{subfigure}
    \caption{Taylor--Green vortex test for different discretizations: black lines are used for \ECb, red for \ECs, green for \ECf, blue \ECw. On the left-hand side, global quantities are evaluated at each time step of the simulation; on the right-hand side, they are sampled every $200$ steps. $\text{CFL} = 0.01$ and $M\approx0.08$.}
    \label{fig:TGV}
\end{figure}

\begin{figure}[h!]
     \centering
    \begin{subfigure}[b]{\textwidth}
         \centering
         \includegraphics[width=0.47\textwidth]{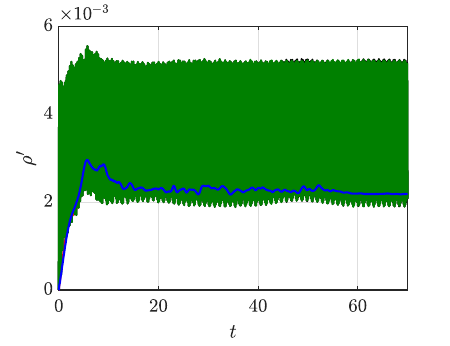}
         \includegraphics[width=0.47\textwidth]{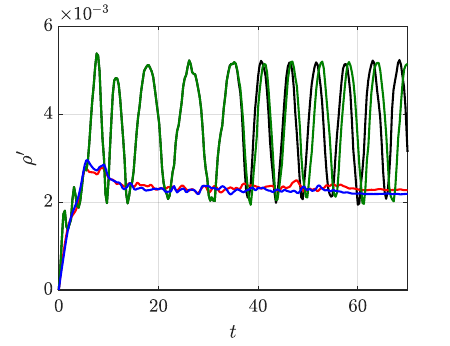}
         \caption{Time evolution of density fluctuations}
         \label{fig:TGV_fluct_rho}
     \end{subfigure}
    \begin{subfigure}[b]{\textwidth}
         \centering
         \includegraphics[width=0.47\textwidth]{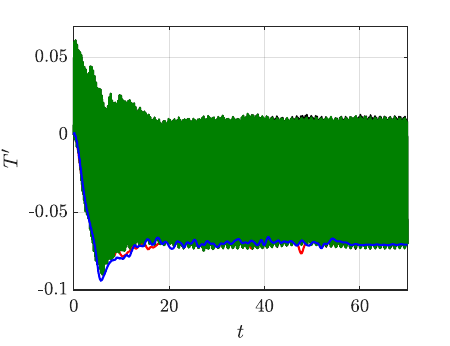}
         \includegraphics[width=0.47\textwidth]{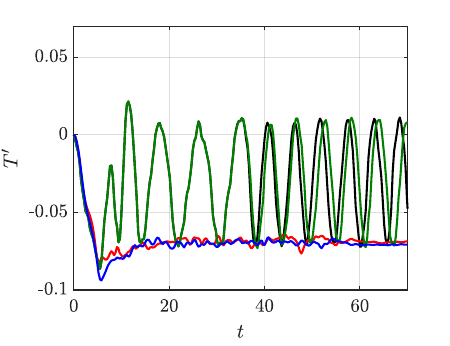}
         \caption{Time evolution of temperature fluctuations}
         \label{fig:TGV_fluct_T}
     \end{subfigure}
    \caption{Taylor--Green vortex test for different discretizations: black lines are used for \ECb, red for \ECs, green for \ECf, blue \ECw. On the left-hand side, fluctuation are evaluated at each time step of the simulation; on the right-hand side, they are sampled every $200$ steps. $\text{CFL} = 0.01$ and $M\approx0.08$.}
    \label{fig:TGV_fluct}
\end{figure}

For this case, the domain has been discretized using $32\times32\times32$ nodes; the final time of the simulation is $T=70$ and the time step size adapts to keep $\CFL = 0.01$.
Entropy is conserved for the test case (Fig.~\ref{fig:TGV_entropy}), but, while for the central schemes \ECw and \ECs the error is down to $10^{-14}$, for the asymmetric schemes \ECb and \ECf there is a drift in the integral value. 
This accumulation of errors is not observed in the conservation of the primary variables density, momentum and total energy (not shown here).
On the other hand, the global value of kinetic energy presents clear oscillations for \ECb and \ECf schemes, which is likely connected to the first order discretization of the pressure term that governs the exchange between kinetic and internal energy. Fig.~\ref{fig:TGV_kinetic} shows the evolution of global kinetic energy considering the value at each time step of the simulation (on the left) and, for additional clarity, sampling every $200$ steps (on the right)\footnote{The under-sampled plots suggest a leading behavior for the \ECf scheme after awhile; however, this is an artifact of the chosen sampling rate.}.
A similar behavior is also displayed for the evolution of density and temperature fluctuation $\rho'$ and $T'$ in Fig.~\ref{fig:TGV_fluct}, with $\phi' = \sqrt{\frac{1}{N}\sum_{i=1}^N (\phi_i - \phi_\mu)^2}$ and $\phi_\mu = \frac{1}{N} \sum_{i=1}^N \phi_i$ is the average value of $\phi$ in the domain with $N$ being the number of nodes. 
However, despite the oscillations, none of the schemes yield an unbounded growth of the thermodynamic fluctuations.
This is a desirable outcome since, after an initial transient, the flow should behave like inviscid isotropic homogeneous turbulence~\cite{Pirozzoli_JCP_2010, Coppola_JCP_2019, Honein_JCP_2004}.
This successful result should not to be taken for granted, as many discretizations fail this test and present uncontrolled production in the fluctuations~\cite{Kuya_JCP_2018,DeMichele_C&F_2023}.

%-----------
\subsubsection{Compressible case\texorpdfstring{ ($M_0=0.4$)}{}}

\begin{figure}[ph!]
     \centering
    \begin{subfigure}[b]{0.47\textwidth}
         \centering
         \includegraphics[width=\textwidth]{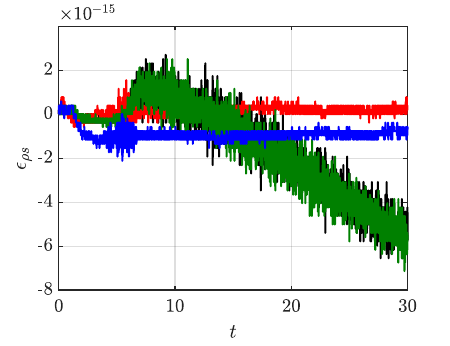}
         \caption{Time evolution of entropy integral}
         \label{fig:TGV_M04_entropy}
     \end{subfigure}
    \begin{subfigure}[b]{0.47\textwidth}
         \centering
         \includegraphics[width=\textwidth]{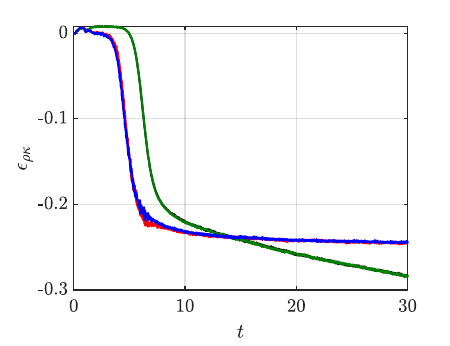}
         \caption{Time evolution of kinetic energy integral}
         \label{fig:TGV_M04_kinetic}
     \end{subfigure}
    \caption{Taylor--Green vortex test for different discretizations: black lines are used for \ECb, red for \ECs, green for \ECf, blue \ECw. Global quantities are evaluated every $2000$ time steps of the simulation. $\text{CFL} = 0.001$ and $M=0.4$. }
    \label{fig:TGV_M04}
% \end{figure}
\vspace{2mm}
% \begin{figure}[h!]
     \centering
    \begin{subfigure}[b]{0.47\textwidth}
         \centering
         \includegraphics[width=\textwidth]{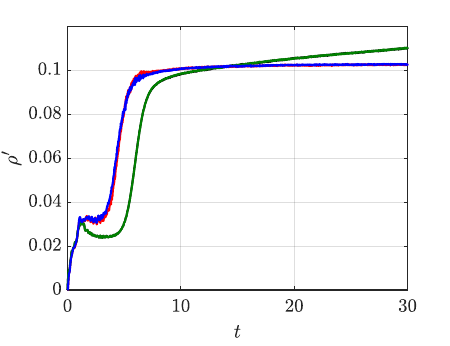}
         \caption{Time evolution of density fluctuations}
         \label{fig:TGV_M04_fluct_rho}
     \end{subfigure}
    \begin{subfigure}[b]{0.47\textwidth}
         \centering
         \includegraphics[width=\textwidth]{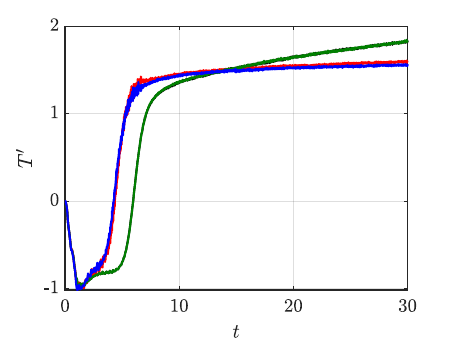}
         \caption{Time evolution of temperature fluctuations}
         \label{fig:TGV_M04_fluct_T}
     \end{subfigure}
    \caption{Taylor--Green vortex test for different discretizations: black lines are used for \ECb, red for \ECs, green for \ECf, blue \ECw. Fluctuations are sampled every $2000$ steps. $\text{CFL} = 0.001$ and $M=0.4$. 
    }
    \label{fig:TGV_M04_fluct}
\end{figure}

For this simulation a $64\times64\times64$ discretization of the domain has been used. A smaller valued time step size has been chosen to obtain $\CFL =0.001$ and prevent errors of the time integrator on entropy conservation. The final time of the simulation is $T=30$.
Even with non-negligible compressibility effects, entropy is confirmed to be conserved by the schemes tested (Fig.~\ref{fig:TGV_M04_entropy}). Similarly to the previous near-incompressible case, a drift is also present here for the asymmetric schemes \ECb and \ECf. However, the oscillations on kinetic energy evolution that were observed in the previous test case are no longer noticeable, even in the absence of time sampling.
This could possibly be due to the increased exchanges of energy between kinetic and internal energies, associated with the pressure-work terms.
However, unlike the nearly incompressible case, for the asymmetric schemes the value of global kinetic energy does not seem to stabilize at the end time of the simulation, but continues to change slightly (Fig.~\ref{fig:TGV_M04_kinetic}).
Similar findings can be found by studying the fluctuations (Fig.~\ref{fig:TGV_M04_fluct}), which also show a drift for \ECb and \ECf schemes but no oscillations.

%
%----
\section{Conclusions} \label{sec:6}

The current work introduces a family of entropy conserving schemes for a calorically perfect single-component gas.
The methods have the additional properties of also being kinetic-energy-preserving and pressure-equilibrium-preserving.
These schemes are, to the best of our knowledge, the first finite-difference discretizations to be locally conservative in both the primary quantities (mass, momentum, total energy) as well as the discrete entropy dynamics.
Their ability to attain exact discrete conservation is confirmed with a variety of test cases in one-, two- and three-dimensions; meanwhile, the propagation of a density wave is also simulated to test the PEP property of the schemes.

The new framework introduces and leverages FD representations of the logarithmic mean, which is commonly employed in EC numerical fluxes. 
This equivalence is just one instance of a more general link between linear and nonlinear two-point interpolants and FD forms involving the ratio of differentials. 
Some of the most commonly used means (e.g., arithmetic, geometric, harmonic) are also provided in this paper.
While the schemes presented in this work are initially introduced in the context of finite differencing, they are also locally conservative due to their SBP property and therefore admit a local flux form that can be used in a FV or FE setting.
Even though we considered uniform grids herein, extension to arbitrary nodal distributions is also possible as long as a diagonal-norm SBP property is available.
The corresponding EC property of the methods is based on point-wise cancellations of spurious volumetric terms that would otherwise be induced in the discrete entropy equation.
The proposed FD methods furthermore are expected to work on general triangle/tetrahedral grids via a recent extension of the SBP framework to tensor-product spectral-element operators \cite{Montoya:2024,Worku:2025}.
In addition, while the current results consider single-block periodic domains, incorporating entropy-stable interface conditions for boundaries and inter-block coupling is possible via Simultaneous Approximation Terms \cite{Zingg:2014,Svard:2021}, as is typically done with SBP-compatible nonlinear flux methods.
Appendix \ref{sec:app_b} furthermore presents the current methods in a Hadamard product form, which is in line with how many modern structure-preserving methods are written for general use, such as with high-order discontinuous Galerkin discretizations \cite{Ranocha:2023}.

Different avenues for future development remain.
First, the schemes presented in this paper admit the possibility of using non-symmetric EC fluxes; such biased forms could be useful in the context of reducing oscillations around sharp gradient fields, and exploring this topic could be the subject of future research. In the meantime, Appendix \ref{sec:app_c} provides a basic coupling of the current methods with entropy-stable artificial dissipation for reducing numerical noise and addressing flow discontinuities.
Secondly, the current methods' extension to high-order stems naturally from employing high-order finite difference stencils to the differential definitions of the specialized averages; however, this will require extending the two-point algorithm for logarithmic means \cite{Ismail_JCP_2009} to a multi-point format that is capable of 1) effectively treating instances where the quotient formula becomes singular and 2) enforcing positivity (i.e., boundedness) of the result.
 In the meantime, however, Appendix \ref{sec:app_d} provides an alternate error-reducing treatment based on flux differencing for the narrow-width biased methods.
Finally, the prospect of extending the current framework to general equations of state per the recent work of \citet{Aiello_JCP_2025,Aiello_ArXiv_2025} is to be explored.

\FloatBarrier
%----

% \newpage
\appendix

%-----------

\section{Demonstrating entropy conservation in SBP matrix-vector form}\label{sec:app_a} 

The demonstrations of entropy conservation from Section \ref{sec:4} are expressed here in terms of matrix-vector manipulations, employing SBP operators.
Note that the following applies to general nodal SBP arrangements, which, besides finite differencing, also includes element-based methods such as discontinuous Galerkin and flux reconstruction discretizations \cite{Gassner:2013,Cicchino:2022}.
A brief introduction to the relevant mechanics of finite difference SBP operators is offered below.
Readers are referred to \cite{Zingg:2014,DRF:2014} for further details.

The finite difference SBP operators considered herein have the following decompositions, presented here in one dimension for simplicity 
\begin{equation}
\begin{array}{c c c c c}
    \bmD_x^\pm = \bmH_x^{-1} \bmQ^\pm & \bmQ^\pm + (\bmQ^\mp)^T = \bmB & \bmH_x = \bmH_x^T > 0 & {\bf a}^T\bmB {\bf b} = (a_rb_r - a_\ell b_\ell)  .
\end{array}
\end{equation}
Each of the above operators carries with it an interpretable role. % that is essential to mimicking integration by parts. 
The symmetric positive definite matrix $\bmH_x$ constitutes a quadrature rule \cite{DRF:2014,Hicken:2013} such that discrete monomials of degree $p$ are accurately integrated (i.e., ${\bf 1}^T\bmH_x{\bf x}^p \approx \frac{1}{p+1}[x_r^{p+1} - x_\ell^{p+1}]$, where $x_{\ell/r}$ is the data at the left/right boundary).
In addition, $\bmB$ is a symmetric and potentially non-diagonal operator that approximates monomial data at the boundaries (i.e., surface integrals in multi-dimensions) such that $({\bf x}^p)^T\bmB{\bf y}^q = (x_r^p y_r^q) - (x_\ell^p y_\ell^q)$; its accuracy may generally be subject to interpolation error except for special instances.
The specialized composition of the SBP derivative operator permits a discrete analogue to integration-by-parts.
Denoting diagonal matrices with brackets (e.g., $[a] \triangleq [\text{diag}\{{\bf a} \}]$) and assuming a diagonal norm in order to leverage commutation, then split forms based on SBP operators can be shown to be discretely conservative 
\begin{eqnarray}
\int_\Omega [a \delta_x b + b \delta_x a]   & \to & {\bf 1}^T \bmH_x\left([a] \bmD^\pm_x[b]{\bf 1}  + [b] \bmD^\mp_x [a]{\bf 1} \right)\\
& = & {\bf 1}^T \left([a] \bmQ^\pm[b] + [b]\bmQ^\mp[a]\right){\bf 1} \nonumber \\
&=& {\bf 1}^T [a]\left(\bmQ^\pm + (\bmQ^\mp)^T\right)[b]{\bf 1} \nonumber \\
&=& {\bf a}^T \bmB{\bf b} \nonumber \\
&=& \underbrace{a_r b_r - a_\ell b_\ell}_{\sim \ (ab)|_{\partial \Omega}} .\nonumber
\end{eqnarray}
As alluded to above, the SBP operator telescopes. 
\citet{Fisher_JCP_2013a} show that such quadratic split forms discretized with diagonal norm SBP operators on nodal points $x_m$ can furthermore be re-written in a flux difference form, $\bmH^{-1}_x \Delta \ \bar{\bf f}$, with the high order fluxes, $\bar{\bf f}$, being composed at intermediate interface locations $x_{m \pm 1/2} \in [x_m,x_{m+1}]$.
Therefore, such split formulations are in fact also locally conservative with respect to a special averaging of the variables, and they adhere to a precise discrete product rule.

\begin{eqnarray*}
    \boldsymbol{\mathcal{C}}_{\rho s} \big|_{\partial \Omega} = \int_\Omega \boldsymbol{\mathcal{C}}_{\rho s} &=&   \int_\Omega \left[\left[s - \gamma c_v \right] \cdot \boldsymbol{\mathcal{C}}_{\rho} + \frac{1}{T} \cdot \left(\boldsymbol{\mathcal{C}}_{\rho e} + \boldsymbol{\mathcal{P}}_{\rho e} \right) \right] \\
       &=& \underbrace{c_v \int_\Omega \left[ \frac{1}{e} \cdot  \boldsymbol{\mathcal{P}}_{\rho e} - (\gamma-1) \log \frac{\rho}{\rho_\text{ref}} \cdot \boldsymbol{\mathcal{C}}_\rho\right]}_{\color{blue}(1)} 
 + \underbrace{c_v\int_\Omega \left[ \left(\log \frac{e}{e_\text{ref}} - \gamma \right) \cdot \boldsymbol{\mathcal{C}}_\rho + \frac{1}{e} \cdot \boldsymbol{\mathcal{C}}_{\rho e}\right]}_{\color{red}(2)}  .
\end{eqnarray*}
Now, we turn to demonstrating the entropy conservation of the FD-EC schemes by employing the aforementioned SBP machinery.
Equations~\eqref{eq:FD-ECTerm1}--\eqref{eq:FD-ECTerm2} in Section~\ref{sec:4} employ local stencil notation in order to motivate the choice of filtered variables $\tilde{\rho}$ and $\tilde{e}$ that are necessary for canceling the spurious volumetric terms {\color{blue}(1)} and {\color{red}(2)} that arise in the discrete entropy equation repeated above.
Here matrix-vector notation is used to convey this  below in  Eqs.~\eqref{eq:FD-ECTerm1_SBP} and \eqref{eq:FD-ECTerm2_SBP}, assuming diagonal-norm SBP differencing operators.
As before, we consider right-biased differencing in one-dimension as an example.

\begin{eqnarray}
&& \underbrace{c_v \int_\Omega \left[ \frac{1}{e} \cdot  \boldsymbol{\mathcal{P}}^{(+)}_{\rho e} - (\gamma-1) \log \frac{\rho}{\rho_\text{ref}} \cdot \boldsymbol{\mathcal{C}}^{(+)}_\rho\right]}_{\color{blue}(1)}: \nonumber \\
 & \to&  c_v(\gamma-1) \cdot \overbrace{{\bf 1}^T\bmH_x \left([\rho] \bmD_x^+[u]{\bf 1} - [\log(\rho/\rho_\text{ref}] \bmD_x^+[\tilde{\rho} u]{\bf 1} \right)}^{\sim \  \int_\Omega \left[ \rho \cdot \delta_x^+ u -  \log \frac{\rho}{\rho_\text{ref}} \cdot \delta_x^+ (\tilde{\rho} u)  \right]} \label{eq:FD-ECTerm1_SBP} \\
 &= &  c_v(\gamma-1) \cdot {\bf 1}^T\left([\rho] (B - (\bmQ^-)^T)[u] - [\log(\rho/\rho_\text{ref}] (B - (\bmQ^-)^T)[\tilde{\rho} u] \right){\bf 1} \nonumber  \\
&=&  c_v(\gamma-1) \cdot \left({\bf 1}^T\left([\rho] B [u] - [\log(\rho/\rho_\text{ref}]B[\tilde{\rho} u] \right){\bf 1}  -  {\bf 1}^T[u]\left(\bmQ_x^-[\rho] - [\tilde{\rho}] \bmQ_x^- [\log(\rho/\rho_\text{ref}] \right) {\bf 1} \right) \nonumber  \\ \nonumber \\
&=&  c_v(\gamma-1) \cdot \left({\bf 1}^T\left([\rho] B [u] - [\log(\rho/\rho_\text{ref}]B[\tilde{\rho} u] \right){\bf 1}  -  \overbrace{{\bf 1}^T[u]H_x\underbrace{\left(\bmD_x^-[\rho] - [\tilde{\rho}] \bmD_x^- [\log(\rho/\rho_\text{ref}] \right) {\bf 1}}_{={\bf 0} \ \text{assuming} \ [\tilde{\rho}] \ \triangleq \ \bmD_x^-[\rho][\bmD_x^- [\log(\rho/\rho_\text{ref}]{\bf 1}]^{-1}}}^{ \sim \ \int_\Omega u \left[ \delta_x^- \rho -  \tilde{\rho} \cdot  \delta_x^-\left( \log \frac{\rho}{\rho_\text{ref}} \right) \right]} \right) \nonumber  \\ \nonumber \\
&=&  \underbrace{ c_v(\gamma-1) \cdot \left(\left(\rho_r u_r - (\log(\rho/\rho_\text{ref})_r (\tilde{\rho} u)_r \right) - \left(\rho_\ell u_\ell - (\log(\rho/\rho_\text{ref})_\ell(\tilde{\rho} u)_\ell \right) \right)}_{\sim \ c_v(\gamma-1) \left[ \rho u - \left( \log \frac{\rho}{\rho_\text{ref}}\right) \cdot \tilde{\rho} u \right] \big|_{\partial \Omega}} \nonumber
%
% && - c_v(\gamma-1) \int_\Omega u \underbrace{\left[ \delta_x^- \rho -  \tilde{\rho} \cdot  \delta_x^-\left( \log \frac{\rho}{\rho_r} \right) \right]}_{\text{need} \ =0} \label{eq:FD-ECTerm1} \\ \nonumber \\
%
\end{eqnarray}
\begin{eqnarray}
    &&\underbrace{c_v\int_\Omega \left[ \left(\log \frac{e}{e_\text{ref}} - \gamma \right) \cdot \boldsymbol{\mathcal{C}}^{(+)}_\rho + \frac{1}{e} \cdot \boldsymbol{\mathcal{C}}^{(+)}_{\rho e}\right]}_{\color{red}(2)}:  \nonumber \\
&\to & c_v \cdot \overbrace{{\bf 1}^T \bmH_x\left([\log(e/e_\text{ref}) - \gamma] \bmD^+_x[\tilde{\rho}u]{\bf 1} + [e^{-1}] \bmD^+_x[\tilde{\rho} u \tilde{e}]{\bf 1} \right)}^{\sim \ \int_\Omega \left[ \left(\log \frac{e}{e_\text{ref}} - \gamma \right) \cdot \delta_x^+ (\tilde{\rho} u )  + e^{-1} \cdot \delta_x^+ (\tilde{\rho} u \tilde{e}) \right]} \label{eq:FD-ECTerm2_SBP} \\
&=& c_v \cdot {\bf 1}^T \left([\log(e/e_\text{ref}) - \gamma] (\bmB - (\bmQ^-)^T)[\tilde{\rho}u]{\bf 1} + [e^{-1}] (\bmB - (\bmQ^-)^T)[\tilde{\rho} u \tilde{e}]{\bf 1} \right) \nonumber \\
&=& c_v \cdot \left({\bf 1}^T \left([\log(e/e_\text{ref}) - \gamma] \bmB[\tilde{\rho}u]{\bf 1} + [e^{-1}] \bmB[\tilde{\rho} u \tilde{e}]{\bf 1} \right) - {\bf 1}^T\left([\tilde{\rho}u] \bmQ^- [\log(e/e_\text{ref}) - \gamma]{\bf 1} + [\tilde{\rho} u \tilde{e}] \bmQ^- [e^{-1}]{\bf 1} \right)\right) \nonumber \\
&=& c_v \cdot \left({\bf 1}^T \left([\log(e/e_\text{ref}) - \gamma] \bmB[\tilde{\rho}u]{\bf 1} + [e^{-1}] \bmB[\tilde{\rho} u \tilde{e}]{\bf 1} \right) - \overbrace{{\bf 1}^T\bmH_x[\tilde{\rho}u]\underbrace{\left( \bmD^-_x [-\log(e^{-1}/e^{-1}_\text{ref})]{\bf 1} + [ \tilde{e}] \bmD^-_x [e^{-1}]{\bf 1} \right)}_{= {\bf 0} \ \text{assuming} \ [\tilde{e}] \ \triangleq \ \bmD^-_x [\log(e^{-1}/e^{-1}_\text{ref})][\bmD^-_x [e^{-1}]]^{-1}}}^{\sim \ \int_\Omega \tilde{\rho} u \left[ -\delta_x^- \left( \log \frac{e^{-1}}{e_\text{ref}^{-1}} \right)  +   \tilde{e} \cdot \delta_x^- e^{-1} \right]}\right) \nonumber \\
&=&  \underbrace{c_v \cdot \left(\left((\log(e_r/e_\text{ref}) - \gamma) \cdot (\tilde{\rho} u)_r +(e^{-1})_r (\tilde{\rho}u\tilde{e})_r\right) - \left((\log(e_\ell/e_\text{ref}) - \gamma) \cdot (\tilde{\rho} u)_\ell +(e^{-1})_\ell (\tilde{\rho}u\tilde{e})_\ell\right)\right)}_{\sim \ c_v \left[ \left(\log \frac{e}{e_\text{ref}} - \gamma \right) \cdot  (\tilde{\rho} u )  + e^{-1} \cdot (\tilde{\rho} u \tilde{e}) \right] \big|_{\partial \Omega}} \nonumber
\end{eqnarray}

As in Section \ref{sec:4}, we see that properly defining the filtered density and internal energy variables allows one to cancel out spurious volumetric terms in a point-wise fashion. 
Note that in practice, one solves for these filtered variables as local inversions.
For example, with respect to the density, one solves the following relation:
\begin{equation}
    [\tilde{\rho}][\bmD_x^\pm[\log(\rho/\rho_\text{ref})]{\bf 1}] = [\bmD_x^\pm[\rho]{\bf 1}] \quad \text{or equivalently} \quad  [\bmD_x^\pm[\log(\rho/\rho_\text{ref})]{\bf 1}] \ {\tilde{\boldsymbol \rho}} = \bmD_x^\pm[\rho]{\bf 1},
\end{equation}
The above assumes that 
$[\bmD_x^\pm[\log(\rho/\rho_\text{ref})]{\bf 1}]$ is invertible.
In cases where this matrix is ill-conditioned (e.g., due to near-constant fields of density which would yield zero entries), then asymptotic estimates are typically employed for determining the filtered quantity.
While such procedures are well understood of two-point differences \cite{Tamaki_JCP_2022,DeMichele_C&F_2023,Kawai:2025}, additional development is required for supporting multi-point differences.

%----------
\section{Expressing the methods in Hadamard product form} \label{sec:app_b}

In the current section, we present the FD-EC schemes in the Hadamard form in order to further generalize their applicability.
While the discretizations thus far have been presented in terms of split-form differencing, an alternate presentation is possible with Hadamard products formed between a matrix $\bmD$ which consists of the difference stencils in space and a flux matrix $F$ that is comprised of two-point flux evaluations.
Such implementations are typically referred to as flux differencing, which is also often used in modern algorithms for structure-preserving methods for high-order discontinuous Galerkin (DG) discretizations~\cite{Ranocha:2023}.
Conveniently, the formalism allows for a flexible substitution of different splittings simply by redefining the flux matrix\footnote{The efficiency of a flux differencing implementation compared to a finite difference algorithm will depend on the specific split form.}.

The two-point fluxes associated with split forms employing non-biased difference operators (e.g., central summation-by-parts operators) are known~\cite{Pirozzoli_JCP_2010,Fisher_JCP_2013b,Coppola_JCP_2023}; however, the extension to split forms based on biased difference operators has not been established to the authors' knowledge and is presented here.
To do this, we leverage the fact that biased schemes may be separated into a central portion and a dissipation portion, as previously shown in Eqs~\eqref{eq:FD-EC_Res_c} and \eqref{eq:FD-EC_Res_AD}. 

Considering the definition of the Hadamard product of two matrices,
\begin{equation}
C = A \circ B \ \to \ C_{ij} = A_{ij} \cdot B_{ij}  \ ,
\end{equation}
where the Hadamard product yields an entry-wise multiplication of matrix components.
We then seek a Hadamard description of the general split-form FD-EC scheme in the direction $x_n$ as
\begin{eqnarray}
    \mathcal{R}_n^{(\omega)} =\left(\mathcal{R}_n^{(\omega),c} + 
\mathcal{R}_n^{(\omega),AD}\right) &=& -\left(\bmD_{x_n}^c \circ 2F_n^{(\omega),c} + \bmD_{x_n}^{AD} \circ 2F_n^{(\omega),AD}\right){\bf 1} \quad \text{with} \ \omega \in [-1,1] \label{eq:fluxdiff_biasedsplit} \\ % \ \text{with} \ D_{x_n}^c = \frac{1}{2}(D + D^T)
&=& - \left(\left(\frac{\bmD_{x_n}^+ + \bmD_{x_n}^-}{2}\right) \circ 2F_n^{(\omega),c} + \left(\frac{\bmD_{x_n}^+ - \bmD_{x_n}^-}{2}\right) \circ 2F_n^{(\omega),AD} \right){\bf 1} \nonumber \\
&=&- \frac{1}{2}\left(\bmD_{x_n}^+ \circ 2\left(F_n^{(\omega),c}  +F_n^{(\omega),AD}\right) + \bmD_{x_n}^- \circ 2\left(F_n^{(\omega),c} - F_n^{(\omega),AD} \right) \right){\bf 1} \nonumber \\
&=&- \frac{1}{2}\left(\bmD_{x_n}^+ \circ 2F_n^{(\omega),+}   + \bmD_{x_n}^- \circ 2F_n^{(\omega),-} \right){\bf 1} \nonumber 
\end{eqnarray}
where the central and artificial dissipation components of the overall operator are based on the symmetric and skew-symmetric portions of the biased stencil per Eqs~\eqref{eq:cent_dp_rel} and \eqref{eq:fd_bias}.
Equation~\eqref{eq:fluxdiff_biasedsplit} provides different factorizations for how to execute the flux differencing.
Note that the above notation implies an equation-by-equation composition (e.g., $\mathcal{R}_n^{(\omega)} \equiv [\mathcal{R}_{n,\rho}^{(\omega)} \ \mathcal{R}_{n,\rho u_i}^{(\omega)} \ \mathcal{R}_{n,\rho E}^{(\omega)}]^T$ and $F_n^{(\omega),\pm} \equiv [F_{n,\rho}^{(\omega),\pm} \  F_{n,\rho u_i}^{(\omega),\pm} \ F_{n,\rho E}^{(\omega),\pm}]^T$ such that $\mathcal{R}_{n,\rho}^{(\omega)} = - \left(\bmD_{x_n}^+ \circ 2F_{n,\rho}^{(\omega),+}   + \bmD_{x_n}^- \circ 2F_{n,\rho}^{(\omega),-} \right){\bf 1}$, etc.).

Inspecting the respective ``central" and ``dissipation" portions from Eqs.~\eqref{eq:FD-EC_Res_c} and \eqref{eq:FD-EC_Res_AD}, one identifies the following two-point flux definitions to be used within the Hadamard formalism for the parameterized FD-EC schemes:
\begin{eqnarray}
    F_n^{(\omega),c}(q_i,q_{i+k}) &=& \left[\begin{array}{l} 
    \frac{(\tilde{\rho}^{(-\omega)} u_n)_{i} + (\tilde{\rho}^{(-\omega)} u_n)_{i+k}}{2} \\ \\
    \frac{(\tilde{\rho}^{(-\omega)} u_n)_{i+k} + (\tilde{\rho}^{(-\omega)} u_n)_{i+k}}{2}\frac{(u_m)_i + (u_m)_{i+k}}{2} + \frac{p_i + p_{i+k}}{2} \\ \\
    \frac{(\tilde{\rho}^{(-\omega)} u_n\tilde{e}^{(-\omega)})_{i} + (\tilde{\rho}^{(-\omega)} u_n\tilde{e}^{(-\omega)})_{i+k}}{2} \\ + \sum_{m=1}^\text{d}\frac{(\tilde{\rho}^{(-\omega)} u_n u_m)_{i+k}(u_m)_i + (\tilde{\rho}^{(-\omega)} u_n u_m)_{i}(u_m)_{i+k}}{2} \\  + \frac{p_i(u_n)_{i+k} + p_{i+k}(u_n)_i}{2}
    \end{array}\right]  \\ \nonumber  \\
    F_n^{(\omega),AD}(q_i,q_{i+k}) &=& \frac{\omega}{2} \cdot \left[
\begin{array}{l}
[(\tilde{\rho}^{(-\omega)} u_n)_{i+k}-(\tilde{\rho}^{(-\omega)} u_n)_{i}] \\ \\
\frac{(u_m)_i + (u_m)_{i+k}}{2}((\tilde{\rho}^{(-\omega)} u_n)_{i+k} - (\tilde{\rho}^{(-\omega)} u_n)_{i}) - [p_{i+k} - p_i] \\ \\
((\tilde{\rho}^{(-\omega)} u_n\tilde{e}^{(-\omega)})_{i+k}-(\tilde{\rho}^{(-\omega)} u_n \tilde{e}^{(-\omega)})_{i}) \\ + \sum_{m=1}^\text{d}((\tilde{\rho}^{(-\omega)} u_n u_m)_i(u_m)_{i+k} - (\tilde{\rho}^{(-\omega)} u_n u_m)_{i+k}(u_m)_i) \\ + (p_i(u_n)_{i+k} - p_{i+k}(u_n)_i)
    \end{array}
    \right].  \label{eq:F-AD}
\end{eqnarray}
Note that employing the FD-EC method in terms of split differences, rather than in terms of flux differencing based on the Hadamard form, is likely more simple and efficient---although such assessments are beyond the current paper's scope and are not performed here.

The flux differencing associated with the central component of Eq.~\eqref{eq:FD-EC_Res_c} is well understood and is therefore not reviewed herein (e.g., see~\cite{Pirozzoli_JCP_2010,Fisher_JCP_2013b}).
Meanwhile, in order to deduce the flux-differencing associated with the AD component of Eq.~\eqref{eq:FD-EC_Res_AD}, we first consider a basic second-order AD term based on a three-point narrow diffusion stencil of the form:
\begin{eqnarray*}
    a \delta_\text{3pt}^{AD} b - b \delta_\text{3pt}^{AD} a &=& a_i (b_{i+k} - 2 b_i + b_{i-k}) - b_i(a_{i+k} - 2 a_i + a_{i-k}) \\
    &=&(a_ib_{i+k} -  a_{i+k}b_i) - 2(a_i b_i - a_i b_i) + (a_i b_{i-k} -  a_{i-k}b_i) \\
    \to F^{AD}(a_i,b_i,a_{i+k},b_{i+k})   &=& a_ib_{i+k} - a_{i+k}b_i =  -F^{AD}(b_i,a_i,b_{i+k},a_{i+k})
\end{eqnarray*}
The above generalizes for an arbitrary symmetric difference stencil, since these can be decomposed as 
\begin{equation*}
\delta^{AD} a = \sum_k \alpha_k \cdot (a_{i+k} - 2 a_i + a_{i-k} ) \ .
\end{equation*}
The above relations then form the basis for the dissipative matrix $F^{AD}$ proposed in Eq.~\eqref{eq:F-AD}.
%

%-----------

%-----
\section{Incorporating entropy-stable artificial dissipation} \label{sec:app_c}
As previously shown in Section~\ref{sec:Sod}, applying EC schemes in the presence of discontinuities will lead to non-physical results, both in terms of oscillations and in terms of the local entropy dynamics.
While it is not the primary focus of the present work, we will briefly explore the incorporation of entropy stable (ES) regularization in this section, as they are better suited for these applications. Rather than enforcing exact entropy conservation at the discrete level, ES schemes ensure satisfaction of the entropy inequality, by which the mathematical entropy does not increase (i.e. the physical entropy does not decrease).
It is possible to obtain such schemes by introducing an additional specialized dissipation to an otherwise baseline EC scheme.
A wide variety of possibilities for such terms exist, which may further include the use of sensors to locally adjust the amount of regularization for improved accuracy.

In this section we adopt a simple approach, introducing a basic artificial dissipation based on the entropy variables $w$ with the sole aim of demonstrating the feasibility of the concept.
The dissipative term is given by $A_{2p}w$ in which, following the formulation of \citet{Mattsson_JSC_2004}, the operator is defined as
\begin{equation}
    A_{2p} = c_{2p} \cdot (-1)^{p+1} \cdot \bmH^{-1} \tilde{\bmD}_p ^T B \tilde{\bmD}_p.
\end{equation}
This corresponds to a term of order 2$p$, in which $\bmD_p = h^{-p} \tilde{\bmD}_p$ is a consistent approximation of the $p$th derivative, $H$ is the norm matrix or SBP quadrature matrix that scales with grid spacing, and $c_{2p}$ is a scaling coefficient satisfying $\text{sign}\left\{(c_{2p}) \cdot (-1)^{p+1} \right\} \ge 0$.
For the $B$ scaling  matrix, we consider $|\partial f^\text{(inv)}/\partial w| \triangleq (Y^{-1} |\Lambda | Y)$, in which the inviscid flux Jacobian with respect to the entropy variables is diagonalized as $(\partial f^\text{(inv)}/\partial w) =Y^{-1} \Lambda Y $ (see \cite{Merriam:1989a} for the eigen-decomposition) and $|\Lambda|$ is the diagonal matrix containing the absolute values of the eigenvalues. As a result, this constitutes characteristics-based matrix dissipation, which is known to be less diffusive \cite{Chandrashekar_CCP_2013}.
In this work, a second-order non-divided difference operator is employed (written here for a bounded 1D domain):
\begin{align}
\tilde{\bmD}_2 = 
\left[\begin{array}{r r r r r} 
 1 & -2 &  1 \\
 1 & -2 &  1 \\
 & &  \ddots \\
 & &1 & -2 &  1 \\
 &&1 & -2 &  1 
\end{array}\right]
\end{align}
which leads to the final expression for the dissipation term that is appended to the baseline scheme, expressed here in both a matrix-vector and a local modified equation form\footnote{Note that the matrix representation shown in Eq.~\eqref{eq:AD2-ES}, formally, would imply the use of Kronecker products such that the difference operator is applied equivalently to each entropy variable.}:  
\begin{equation}
    d_tq = \dots + \overbrace{\bmH^{-1} \tilde{\bmD}_2^T (Y^{-1} |\Lambda| Y )\tilde{\bmD}_2 w}^{A_4 w} \ \to \ (d_t q_i) \big|_x = \left.\left(\dots + \sum_{j=1}^{(2+\text{dim})} (\Delta x)^3 \cdot \partial_x^2|\partial f_i/ \partial w_j| \partial_x^2w_j  \right) \right|_x \label{eq:AD2-ES}
\end{equation}
Entropy stability is thus automatically recovered since $w^TA_{2p}w \ge 0$ by construction.
Additional tuning of the spectral attenuation of such operators is possible by employing multi-derivative stabilization terms, such as those arising from filter-based dissipation \cite{Edoh:2018,Edoh:2019a}.
Also note that analogous difference operators are available for non-uniform nodal arrangements~\cite{Ranocha:2018,Hicken:2020,Bercik_ArXiv_2025}.
By adding such terms to the EC schemes introduced earlier, we obtain the corresponding ES counterpart, which will be denoted \ESb, \ESf, \ESs, and \ESw in analogy with the respective biasing of the studied FD-EC schemes.

To evaluate the performance of these schemes, we apply them to the 1D Sod shock tube problem, which features discontinuities that can lead to spurious oscillations if not properly treated.
The setup is identical to that described in Section~\ref{sec:Sod}, except that a higher CFL number ($\CFL=0.1$) is used here. This is justified by the fact that the entropy production introduced by the temporal integrator is negligible compared to that introduced by artificial dissipation.

\begin{figure}[h!]
     \centering
    \begin{subfigure}[b]{0.47\textwidth}
         \centering
         \includegraphics[width=\textwidth]{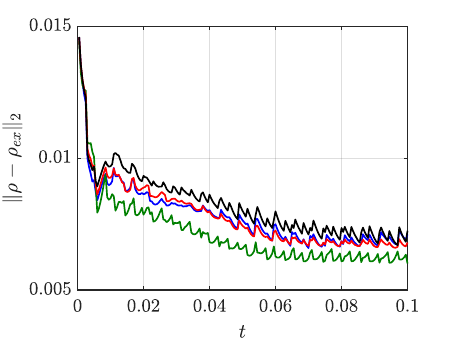}
         \caption{Density}
         \label{fig:Sod_ES_rho_norm}
     \end{subfigure}
    \begin{subfigure}[b]{0.47\textwidth}
         \centering
         \includegraphics[width=\textwidth]{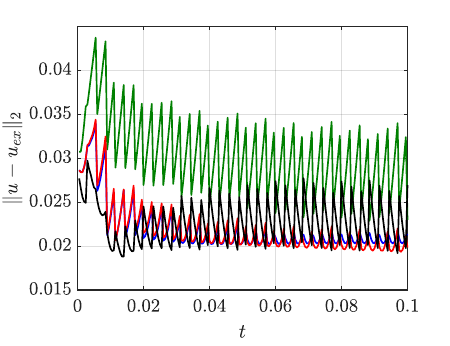}
         \caption{Velocity}
         \label{fig:Sod_ES_u_norm}
     \end{subfigure}
    \begin{subfigure}[b]{0.47\textwidth}
         \centering
         \includegraphics[width=\textwidth]{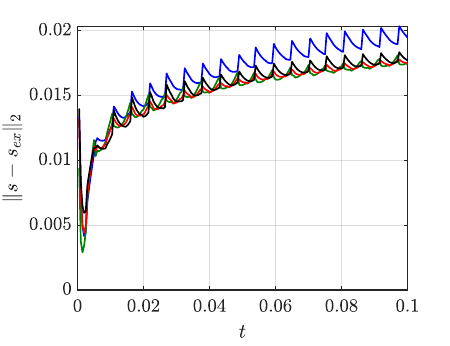}
         \caption{Entropy}
         \label{fig:Sod_ES_s_norm}
     \end{subfigure}
    \begin{subfigure}[b]{0.47\textwidth}
         \centering
         \includegraphics[width=\textwidth]{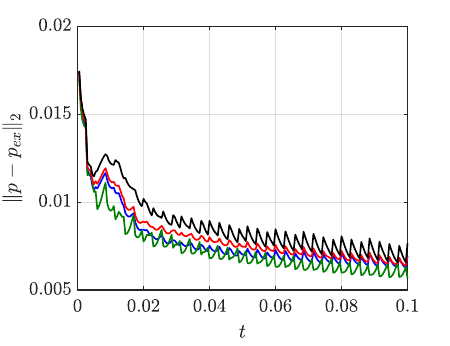}
         \caption{Pressure}
         \label{fig:Sod_ES_T_norm}
     \end{subfigure}
    \caption{Sod test using entropy stable schemes: black lines are used for \ESb, red for \ESs, green for \ESf, blue \ESw.
 $\CFL = 0.1$. Evolution in time
of the $L_2$ norm of the difference between computed solution and analytical one.}\label{fig:Sod_ES_norm}
\end{figure}

\begin{figure}[hp!]
     \centering
    \begin{subfigure}[b]{0.47\textwidth}
         \centering
         \includegraphics[width=\textwidth]{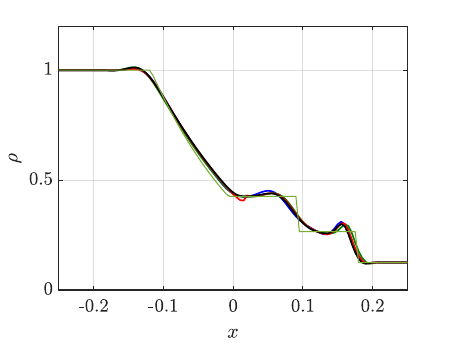}
         \caption{Density}
         \label{fig:Sod_ES_rho}
     \end{subfigure}
    \begin{subfigure}[b]{0.47\textwidth}
         \centering
         \includegraphics[width=\textwidth]{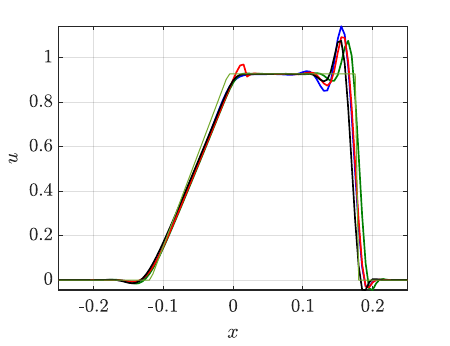}
         \caption{Velocity}
         \label{fig:Sod_ES_u}
     \end{subfigure}
    \begin{subfigure}[b]{0.47\textwidth}
         \centering
         \includegraphics[width=\textwidth]{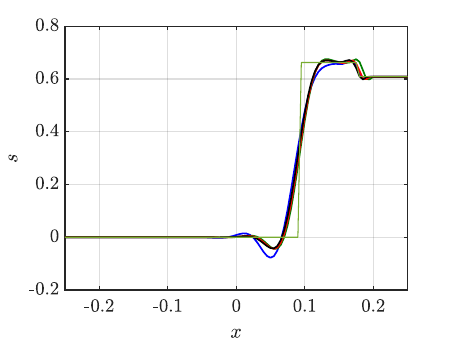}
         \caption{Entropy}
         \label{fig:Sod_ES_s}
     \end{subfigure}
    \begin{subfigure}[b]{0.47\textwidth}
         \centering
         \includegraphics[width=\textwidth]{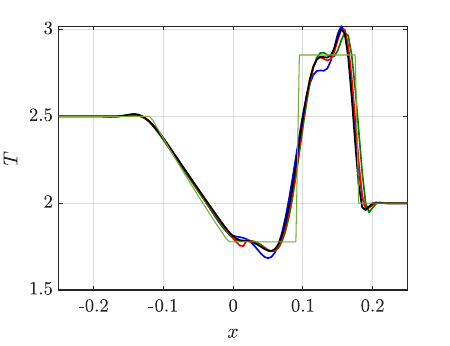}
         \caption{Temperature}
         \label{fig:Sod_ES_T}
     \end{subfigure}
    \begin{subfigure}[b]{0.47\textwidth}
         \centering
         \includegraphics[width=\textwidth]{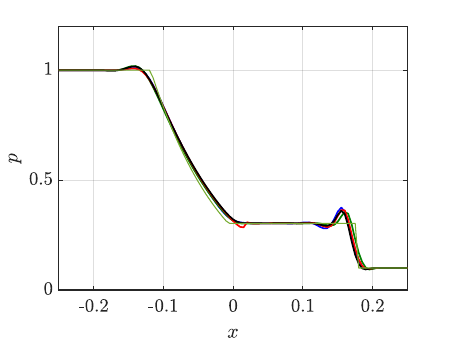}
         \caption{Pressure}
         \label{fig:Sod_ES_p}
     \end{subfigure}
    \begin{subfigure}[b]{0.47\textwidth}
         \centering
         \includegraphics[width=\textwidth]{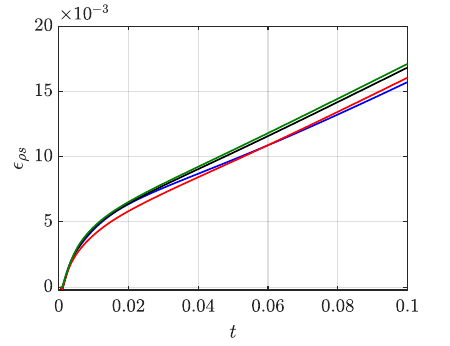}
         \caption{Entropy integral evolution}
         \label{fig:Sod_ES_entropy}
     \end{subfigure}
    \caption{Sod test using entropy stable schemes: black lines are used for \ESb, red for \ESs, green for \ESf, blue \ESw; pale green line is the exact solution. $\CFL = 0.1$, final solutions at $T=0.1$.}
    \label{fig:Sod_ES}
\end{figure}

Figure~\ref{fig:Sod_ES_norm} reports the $L_2$ norm of the error for the various quantities. Improvements are especially visible in the density $\rho$ and pressure $p$ compared to the EC scheme. A noticeable difference emerges when comparing the forward and backward formulations: the \ESf scheme exhibits a larger error in the velocity field compared to the others.
This is in line with the idea that a local choice of the formulation based on the direction of the velocity could be beneficial.

Examining the solution profiles in Fig.~\ref{fig:Sod_ES}, we observe that while some oscillations remain near the discontinuities relative to the exact solution, the ES schemes significantly reduce these artifacts compared to the EC case. Further improvements could potentially be achieved by locally adjusting the dissipation---for example, via scaling, selecting the dissipation order, or switching between scalar and matrix-based dissipation operators.

As expected, all the scheme provide a positive entropy production (see Fig.~\ref{fig:Sod_ES_entropy}) and  demonstrate sufficient accuracy compared to the EC baseline.
For examples, as visible in Fig.~\ref{fig:Sod_ES_s}, all the schemes are now able to reproduce the correct behavior for entropy in the region leading up to shock $x \in [0.1,0.2]$, as opposed to what was happening for the EC versions. 
However, the wide-stencil variant (\ESw) appears to introduce excessive dissipation in certain regions, although this could also potentially be mitigated with appropriate local calibration of the dissipation.
One noteworthy observation is that the \ESs scheme still shows a small spurious peak in pressure and velocity around $x=0$, albeit less pronounced than in the corresponding \ECs solution.

\section{Reducing errors in the biased EC schemes via traditional flux differencing} \label{sec:app_d}

The current FD-EC methods would naturally extend to high-order upon employing high-order derivative formulas within the differencing as well as the FD definitions of the specialize averages. 
However, the latter requires the development of new multi-point algorithms for calculating the mean quantities while  1) appropriately handling the risk of a singularity when evaluating the quotient expressions and 2) enforcing positivity of the output.
Here, in the meantime, we propose a error-reducing extension of the narrow biased FD-EC schemes based on flux differencing.

Per Remark~\ref{remark:2}, the FD-EC methods imply satisfaction of a local entropy conserving flux condition. 
High-order flux reconstructions are then typically viable via combinations of symmetric two-point fluxes \cite{Fisher_JCP_2013b}.
The two-point functions of the biased schemes, however, are no longer symmetric in their arguments (i.e., the order in which the solution vectors $q$ are supplied informs the respective biases of the velocity and the pressure); therefore, this nuance needs to be accounted for within the flux differencing definition as
\begin{equation}
    \frac{\mathcal{F}^{(n)}_{i+1/2} - \mathcal{F}^{(n)}_{i-1/2}}{h_i \cdot \Delta x} = \frac{1}{\Delta x} \left(\sum_{k \le 0} 2d_{i,i+k}^{\text c}  \cdot \mathcal{F}^{(n)}\big|^\omega_{\text{2pt}}(q_{i+k},q_i) + \sum_{k > 0} 2d_{i,i+k}^{\text c}  \cdot \mathcal{F}^{(n)}\big|^\omega_{\text{2pt}}(q_i,q_{i+k}) \right) 
\end{equation}
where 
\begin{equation}
    \mathcal{F}^{(n)}\big|^\omega_{\text{2pt}}(q_a,q_b) = \left[\begin{array}{c} 
    \mathcal{F}^{(n)} _{\rho} \big|^\omega_{\text{2pt}} \\ \\ \mathcal{F}^{(n)} _{\rho u_m} \big|^\omega_{\text{2pt}} \\ \\ \mathcal{F}^{(n)} _{\rho E} \big|^\omega_{\text{2pt}}
    \end{array} \right] =  \left[
    \begin{array}{c}
    \overline{(\rho_a,\rho_b)}^{\log,n} \left((1-\omega) \cdot u_{n,a} + \omega \cdot u_{n,b} \right) \\ \\
    \mathcal{F}^{(n)}_\rho \big|^\omega_{\text{2pt}} \cdot \frac{(u_{m,b} + u_{m,a})}{2} + \delta_{mn} \cdot \left((1-\omega) \cdot p_{b} + \omega  \cdot p_a\right) \\ \\
    \mathcal{F}^{(n)}_\rho \big|^\omega_{\text{2pt}}\cdot \left(\sum_{m=1}^d \frac{u_{m,a} u_{m,b}}{2} + \overline{(e_a,e_{b})}^{H\log,n} \right) \\ + \ (1-\omega) \cdot u_{n,a}p_{b} + \omega \cdot u_{n,b}p_a
    \end{array}
    \right] \ .
\end{equation}
In the above, $d_{i,i+k}^{\text c}$ are the stencil coefficients associated with a \emph{central} SBP operator evaluated at node $i$.
While the fluxes are fully biased with respect to the advective velocity and the pressure, these quantities are \emph{counter-balanced} and therefore the overall method and its fluxes can be interpreted as being ``centered".
As such, a central SBP operator is still appropriate for the prescription of the $d_{i,i+k}^{\text c}$ coefficients.
The multi-point flux associated with the above is (see Eq.~(3.9) in \cite{Fisher_JCP_2013b})
\begin{equation}
    \mathcal{F}^{(n)}\big|^\omega_{\text{multi-pt}} = \sum_{k > 0} \sum_{\ell \le 0} 2d_{i+\ell,i+k}^{\text c}  \cdot \mathcal{F}^{(n)}\big|^\omega_{\text{2pt}}(q_{i+\ell},q_{i+k}) \ .
\end{equation}
Then the Hadamard product form for the residual associated with the above is given as
\begin{equation*}
    \mathcal{R}^{(\omega)} = -(\bmD_{x_n}^c \circ 2F_n^{(\omega)}){\bf 1}
\end{equation*}
with
\begin{equation}
    F_n^{(\omega)}(q_{i+k},q_i) =   \left[
    \begin{array}{c}
    \overline{(\rho_{i+k},\rho_i)}^{\log,n} \left((1-\omega) \cdot u_{n,i+\min[0,k]} + \omega \cdot u_{n,i+\max[0,k]} \right) \\ \\
    \left[\overline{(\rho_{i+k},\rho_i)}^{\log,n} \left((1-\omega) \cdot u_{n,\min[0,k]} + \omega \cdot u_{n,i+\max[0,k]} \right)\right] \cdot \frac{(u_{m,i+k} + u_{m,i})}{2} \\
    \ + \delta_{mn} \cdot \left((1-\omega) \cdot p_{i+\max[0,k]} + \omega  \cdot p_{i+\min[0,k]}\right) \\ \\
    \left[\overline{(\rho_{i+k},\rho_i)}^{\log,n} \left((1-\omega) \cdot u_{n,i+\min[0,k]} + \omega \cdot u_{n,i+\max[0,k]} \right)\right] \cdot \left(\sum_{m=1}^d \frac{u_{m,i+\min[0,k]} u_{m,i+\max[0,k]}}{2} \right. \\
    \left.+ \overline{(e_{i+k},e_i)}^{H\log,n} \right) \\ + \ (1-\omega) \cdot u_{n,i+\min[0,k]}p_{i+\max[0,k]} + \omega \cdot u_{n,i+\max[0,k]}p_{i+\min[0,k]}
    \end{array}
    \right] \ .
\end{equation}

The strict biasing of the velocity and pressure that results from the asymmetric flux function can be expected to limit the above to first order. 
Namely, despite the $d_{i,i+k}$ coefficients  stemming from a high-order central operator, they do not allow for the necessary error cancellations within the velocity and pressure variables, which are fully biased.
Even though formal asymptotic accuracy is not increased, employing such stencils can be made to reduce errors in the solutions.

\begin{figure}[h!]
    \begin{subfigure}[b]{0.47\textwidth}
         \centering
    \includegraphics[width=\textwidth]{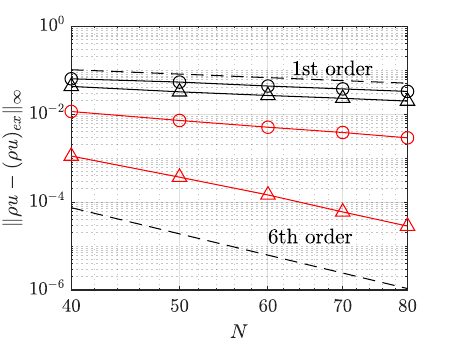}
         \caption{
         Order of accuracy on solution of $\rho u$.
         Circles identify the two-point schemes, triangles the multi-point version.}
         \label{fig:IV_accuracy_6}
     \end{subfigure}
    \begin{subfigure}[b]{0.47\textwidth}
         \centering
         \includegraphics[width=\textwidth]{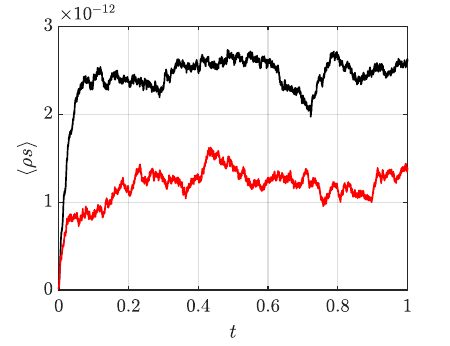}
         \caption{Time evolution of entropy integral}
         \label{fig:IV_entropy_6}
     \end{subfigure}
    \caption{Isentropic vortex test for different discretizations using flux differencing: black lines are used for \ECb and  red for \ECs.
    $\text{CFL} = 0.01$.}
    \label{fig:IV_6}
\end{figure}

\begin{figure}[h!]
     \centering
    \begin{subfigure}[b]{0.47\textwidth}
         \centering
         \includegraphics[width=\textwidth]{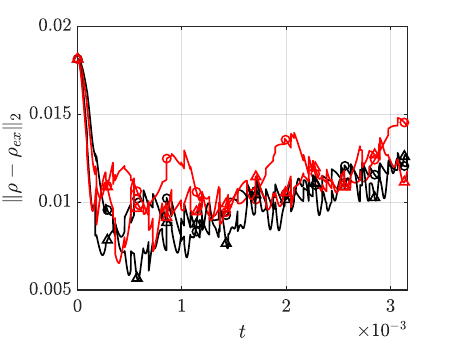}
         \caption{Density}
         \label{fig:Sod_rho_norm_6}
     \end{subfigure}
    \begin{subfigure}[b]{0.47\textwidth}
         \centering
         \includegraphics[width=\textwidth]{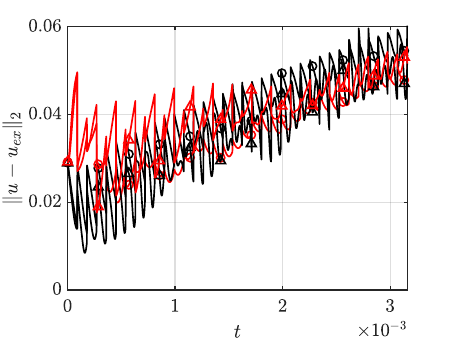}
         \caption{Velocity}
         \label{fig:Sod_u_norm_6}
     \end{subfigure}
    \begin{subfigure}[b]{0.47\textwidth}
         \centering
         \includegraphics[width=\textwidth]{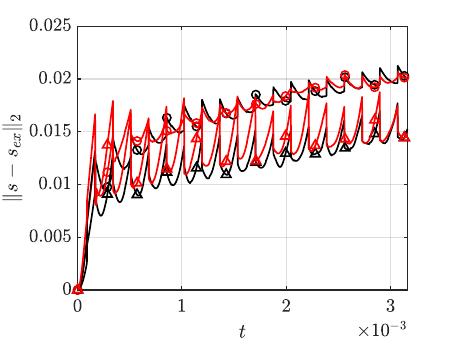}
         \caption{Entropy}
         \label{fig:Sod_s_norm_6}
     \end{subfigure}
    \begin{subfigure}[b]{0.47\textwidth}
         \centering
         \includegraphics[width=\textwidth]{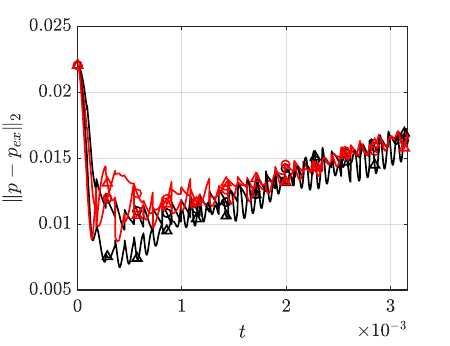}
         \caption{Pressure}
         \label{fig:Sod_T_norm_6}
     \end{subfigure}
    \caption{Sod test for different discretizations: solid black lines are used for \ECb, red for \ECs (note: \ECf would be slightly above the red line of \ECs). $\CFL = 0.01$.  Circles identify the two-point schemes, triangles the multi-point version. Evolution in time
of the $L_2$ norm of the difference between computed solution and analytical one.
}
    \label{fig:Sod_norm_6}
\end{figure}

\begin{figure}[ph!]
     \centering
    \begin{subfigure}[b]{0.47\textwidth}
         \centering
         \includegraphics[width=\textwidth]{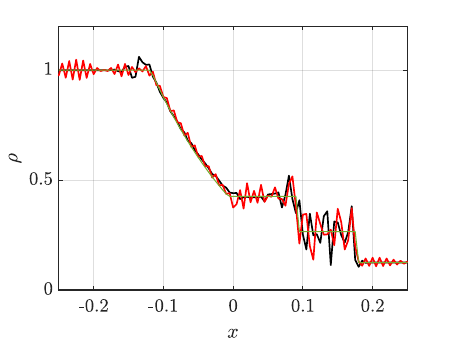}
         \caption{Density}
         \label{fig:Sod_rho_6}
     \end{subfigure}
    \begin{subfigure}[b]{0.47\textwidth}
         \centering
         \includegraphics[width=\textwidth]{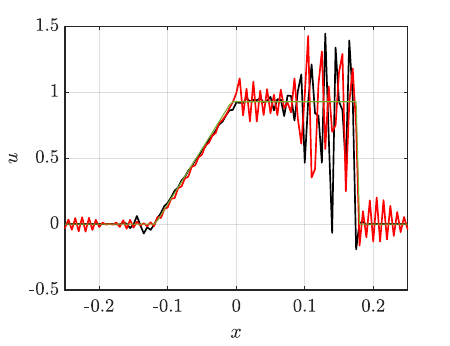}
         \caption{Velocity}
         \label{fig:Sod_u_6}
     \end{subfigure}
    \begin{subfigure}[b]{0.47\textwidth}
         \centering
         \includegraphics[width=\textwidth]{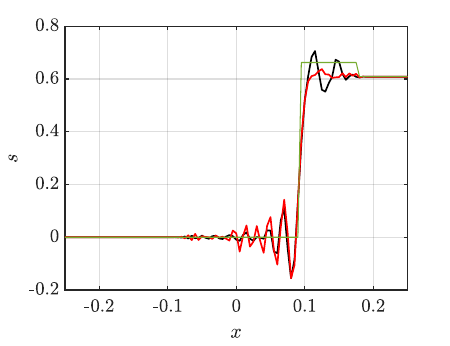}
         \caption{Entropy}
         \label{fig:Sod_s_6}
     \end{subfigure}
    \begin{subfigure}[b]{0.47\textwidth}
         \centering
         \includegraphics[width=\textwidth]{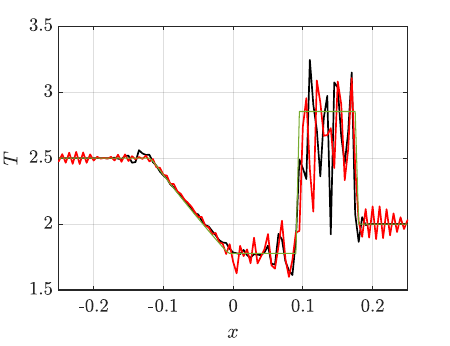}
         \caption{Temperature}
         \label{fig:Sod_T_6}
     \end{subfigure}
    \begin{subfigure}[b]{0.47\textwidth}
         \centering
         \includegraphics[width=\textwidth]{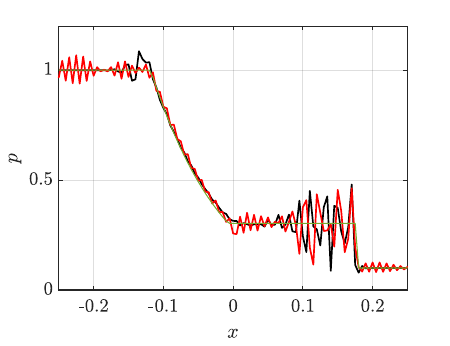}
         \caption{Pressure}
         \label{fig:Sod_p_6}
     \end{subfigure}
    \begin{subfigure}[b]{0.47\textwidth}
         \centering
         \includegraphics[width=\textwidth]{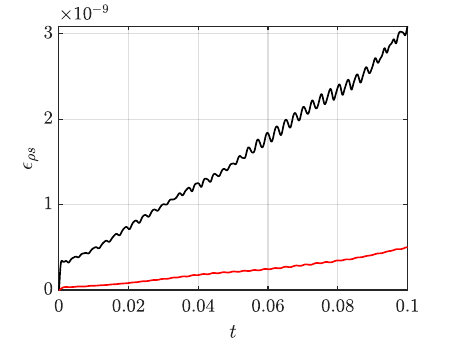}
         \caption{Entropy integral evolution}
         \label{fig:Sod_entropy_6}
     \end{subfigure}
    \caption{Sod test for different discretizations using flux differencing: black lines are used for \ECb and red for \ECs; pale green line is the exact solution. Final solutions at $T=0.1$ based on $\text{CFL} = 0.01$.
    }
    \label{fig:Sod_6}
\end{figure}

Here we revisit the isentropic vortex and Sod shock tube test cases from Section~\ref{sec:5} in order to survey the newly proposed methods; we focus on the backward biased \ECb and symmetric \ECs renditions for brevity.
The long-time entropy conservation of the new formulations are confirmed (see Figs.~\ref{fig:IV_entropy_6} and \ref{fig:Sod_entropy_6}).
Meanwhile, the convergence test on the isentropic vortex (see Fig.~\ref{fig:IV_accuracy_6}) shows that the current multi-point flux-differencing of the biased schemes are still limited to first order; however, the error constant is reduced from the nominal two-point second order method previously shown in Fig.~\ref{fig:IV_accuracy}.
Despite not being formally high-order, the current multi-point flux-differencing extension of the biased schemes is also seen to reduce oscillations in Sod shock tube results compared to the symmetric flux.
In fact, Fig.~\ref{fig:Sod_norm_6} shows that the $L_2$ norm of the error is consistently lower for the biased \ECb scheme in all cases, in spite of the order of accuracy formally being lower. The opposite would be true for the \ECf scheme (not shown here) which would have a slightly larger error for this test case.
The presence of additional non-physical oscillations when the symmetric \ECs scheme is employed can be observed in Fig.~\ref{fig:Sod_6}, which shows the solution at time $T=0.1$.
This property of oscillations reduction therefore motivates the potential utility of such schemes for flows exhibiting sharp gradients, albeit still requiring suitable coupling with entropy stable regularization in order to sufficiently damp large oscillations.
%----
%---
%----
%------
\section*{Acknowledgments}
The author would like to thank the anonymous reviewers for providing helpful comments and suggestions toward the refinement of the current manuscript. CDM and GC acknowledge the CINECA award under the ISCRA initiative, for the availability of high-performance computing resources and support. AE acknowledges funding for this work from the Air Force Office of Scientific Research (AFOSR) (program officers: Drs. Chiping Li, Fariba Farhoo, and Justin Koo) under contract No. 25RQCOR004, as well as Amentum under contract No. FA9300-20-F-9801.
%\vspace{1cm}

\bibliographystyle{model1-num-names}
\bibliography{Biblio_KEP_Compr}

\end{document}